\documentclass[a4paper,twoside,journal]{IEEEtran}
\usepackage{cite}
\ifCLASSINFOpdf
\else
\fi
\usepackage[cmex10]{amsmath}
\usepackage{units}
\usepackage{mathrsfs}
\usepackage{amssymb}
\usepackage{bm}

\usepackage{tikz}
\usetikzlibrary{arrows}

\usepackage{algorithm}

\usepackage{multirow}

\usepackage{algorithmic}
\newtheorem{thm}{Theorem}
\newtheorem{defn}{Definition}
\newtheorem{lem}{Lemma}
\newtheorem{cor}{Corollary}
\newtheorem{rmk}{Remark}

\begin{document}

\title{An Outage Exponent Region based Coded $f$-Matching Framework for Channel Allocation in Multi-carrier Multi-access Channels}


\author{Bo~Bai,~\IEEEmembership{Member,~IEEE,}
        Wei~Chen,~\IEEEmembership{Senior~Member,~IEEE,}\\
        Khaled B. Letaief,~\IEEEmembership{Fellow,~IEEE,}
        and~Zhigang~Cao,~\IEEEmembership{Senior~Member,~IEEE}% <-this % stops a space

\thanks{%Manuscript received MONTH DAY, YEAR; revised MONTH DAY, YEAR; accepted MONTH DAY, YEAR.
This work is supported in part by National Basic Research Program of China (973 Program) under Grants 2013CB336600 and 2012CB316001, RGC under Grant 610311, and Chuanxin Funding. This paper was presented in part at IEEE ICC 2009 and IEEE ICC 2013.}% <-this % stops a space
\thanks{B. Bai, W. Chen, and Z. Cao are with Tsinghua National Laboratory for Information Science and Technology (TNList), and Department of Electronic Engineering, Tsinghua University, Beijing 100084, China (e-mail: eebobai@tsinghua.edu.cn; wchen@tsinghua.edu.cn; czg-dee@tsinghua.edu.cn).}% <-this % stops a space
\thanks{K. B. Letaief is with the Center for Wireless Information Technology, Department of Electronic and Computer Engineering, Hong Kong University of Science and Technology, Kowloon, Hong Kong (e-mail: eekhaled@ust.hk).}% <-this % stops a space
%\thanks{Communicated by XXXX, Associate Editor for Communications.}% <-this % stops a space
%\thanks{Digital Object Identifier }% <-this % stops a space
}


\markboth{IEEE Transactions on Information Theory, vol. XX, no. XX, MONTH YEAR}%
{Bai \MakeLowercase{\textit{et al.}}: Coded $f$-Matching Framework for Channel Allocation in Multi-carrier Multi-access Channels}



\maketitle

\begin{abstract}
  \boldmath
  The multi-carrier multi-access technique is widely adopt in future wireless communication systems, such as IEEE 802.16m and 3GPP LTE-A. The channel resources allocation in multi-carrier multi-access channel, which can greatly improve the system throughput with QoS assurance, thus attracted much attention from both academia and industry. There lacks, however, an analytic framework with a comprehensive performance metric, such that it is difficult to fully exploit the potentials of channel allocation. This paper will propose an analytic coded $f$-matching framework, where the outage exponent region (OER) will be defined as the performance metric. The OER determines the relationship of the outage performance among all of the users in the full SNR range, and converges to the diversity-multiplexing region (DMR) when SNR tends to infinity. To achieve the optimal OER and DMR, the random bipartite graph (RBG) approach, only depending on $\unit[1]{bit}$ CSI, will be proposed to formulate this problem. Based on the RBG formulation, the optimal frequency-domain coding based maximum $f$-matching method is then proposed. By analyzing the combinatorial structure of the RBG based coded $f$-matching with the help of saddle-point approximation, the outage probability of each user, OER, and DMR will be derived in closed-form formulas. It will be shown that all of the users share the total multiplexing gain according to their rate requirements, while achieving the full frequency diversity, i.e., the optimal OER and DMR. Based on the principle of parallel computations, the parallel vertices expansion \& random rotation based Hopcroft-Karp (PVER\textsuperscript{2}HK) algorithm, which enjoys a logarithmic polynomial complexity, will be proposed. The simulation results will not only verify the theoretical derivations, but also show the significant performance gains.
\end{abstract}

% Note that keywords are not normally used for peer review papers.

% \fi
\IEEEpeerreviewmaketitle

\section{Introduction}
\IEEEPARstart{M}{ulti-carrier} technology has shown great potentials in improving the transmission efficiency and reliability in fading channels \cite{Cimini1985}. The multi-carrier based multi-access schemes, such as orthogonal frequency division multi-access (OFDMA) and single-carrier frequency division multi-access (SC-FDMA)\footnote{SC-FDMA can be seen as a coded OFDMA scheme, which is discussed in Section \ref{subsec:coding_schemes}.} etc., have been widely adopt by IEEE 802.16 and 3GPP LTE-A standards \cite{IEEE802.16-2009,3GPPLTE}. Both academia and industry believe that the multi-carrier multi-access technology will be playing a more and more important role in future wireless communication systems \cite{Yang2009}.

It is a key advantage that the multi-carrier multi-access channel provides great flexibility and adaptability in sharing the channel resources among multiple users, because the channel resources are composed by plenty of small \emph{resource blocks} (RBs) in time-frequency domain \cite{Bahai04}. In this context, a RB corresponds to a subcarrier in the duration of a timeslot. Previous works for channel resources allocation can be roughly divided into two categories: RB based scheme, and chunk based scheme. The RB based scheme implies that the based station (BS) directly allocates RBs, which are normally not adjacent in time-frequency domain, to every user. The interleaved subcarrier allocation in IEEE 802.16 and 3GPP LTE-A is a typical RB based allocation scheme, where the uniformly-spaced subcarriers are allocated to each user \cite{IEEE802.16-2009,3GPPLTE}. In \cite{Wong1999}, Wong, Cheng, Letaief, and Murch proposed an OFDMA system and studied the adaptive subcarrier, bit, and power allocation problem. In \cite{Hoo2004}, Hoo, Halder, Tellado, and Cioffi considered allocating subcarriers and power in multi-carrier broadcast channels to maximize the weighted sum rate. In \cite{Song2005}, Song and Li proposed a cross-layer optimization framework for subcarrier and power allocation in OFDM wireless networks. In contrast to RB based schemes, the chunk based scheme means that multiple RBs are bundled together as a chunk. The BS then allocates these chunks to every user. In IEEE 802.16 and 3GPP LTE-A, a typical chunk based allocation scheme is referred to as the localized allocation, where the adjacent subcarriers in the frequency domain are allocated to each user \cite{IEEE802.16-2009,3GPPLTE}. In \cite{Shen2005}, Shen, Li, and Liu studied the performance of a chunk based scheme, where the chunks are allocated to users according to their average SNR within each chunk. Zhu and Wang developed this idea and proposed a joint chunk, power, and bit allocation scheme for OFDMA systems in \cite{Zhu2012}. The basic principle behind these schemes is to maximize the system capacity or some utility function by dynamically allocating RBs/chunks and power under the constraints of total power, fairness, rate, and latency \cite{Stuber05}. The used mathematical tools are mainly based on convex optimization and integral programming \cite{Liu2005,Hanzo06}.

Although many iterative or heuristic algorithms are derived from the aforementioned works to improve the transmission efficiency or enhance the reliability, it is not trivial to conduct a thorough analysis on the fundamental tradeoff between them. Since there lacks an analytic framework with a comprehensive performance metric, the relationship among the efficiency-reliability tradeoffs of different users is also still unknown. Moreover, these proposed allocation schemes usually require perfect channel state information (CSI) at the BS. Based on these considerations, an analytic coded $f$-matching framework is proposed for channel resources allocation problem in multi-carrier multi-access channels. The outage exponent region (OER), which is the region bounded by outage exponents of all the users at given target rates and SNR, is first defined as a fundamental performance metric. In this context, the outage exponent, proposed in our previous work \cite{Bai2011c}, characterizes the relationship among the outage probability, the number of channels, and SNR at the same time. As a matter of fact, the region bounded by diversity-multiplexing tradeoffs (DMTs) of every user, referred to as the diversity-multiplexing region (DMR), is actually the asymptotic OER when SNR tends to infinity. Similar to the capacity region, the proposed OER and DMR comprehensively illustrate the relationship among the achieved outage performances of all the users for a given resources allocation scheme in slow fading multi-access channels. To achieve the optimal OER and DMR, the random bipartite graph (RBG) \cite{Bollobas01} approach, which only depends on $\unit[1]{bit}$ CSI per coherence bandwidth per user, is then applied to formulate the channel resources allocation problem in multi-carrier multi-access channels. The frequency-domain coding based maximum $f$-matching method, which can be applied to both RB and chunk based allocation schemes, is then proposed to allocate the channel resources to all of the users according to their different target rates.

In contrast to previous works, the outage probability, OER, and DMR of the coded $f$-matching framework can be analyzed in closed-form formulas. By applying the saddle-point approximation method, which has a good balance between the approximation accuracy and complexity \cite{Butler2007}, the tight upper bound has been first derived for the outage probability of the optimal coding scheme with $\unit[1]{bit}$ CSI feedback per coherence bandwidth. The combinatorial structure of the RBG based maximum $f$-matching is then studied by using the results in matching theory \cite{Lovasz1986}. Based on these results, the approximation formulas for the outage probability of each user are obtained for RB and chunk based coded $f$-matching schemes in the high and low SNR regimes, respectively. The best achievable OER and DMR for RB and chunk based coded $f$-matching schemes are then obtained accordingly. Although there are serious conflicts in channel resources allocation, the proposed coded $f$-matching framework is still capable of achieving the optimal OER and DMR with multi-user diversity. As a result, all of the users share the total multiplexing gain according to their target rates, while achieving the full frequency diversity simultaneously.

The low-complexity channel resources allocation algorithm and coding scheme are also very important for practical multi-carrier multi-access systems. By applying the principle of parallel computations \cite{Karpinski1998}, we propose the parallel vertices expansion \& random rotation based Hopcroft-Karp (PVER\textsuperscript{2}HK) algorithm. It can be shown that the time complexity is only $\mathcal{O}\left(\log^{2\eta}N\right)$, where $N$ is the number of resource blocks and $\eta$ is a constant. The proposed PVER\textsuperscript{2}HK algorithm is even faster than FFT, $\mathcal{O}\left(N\log N\right)$. To achieve the optimal DMR, two practical DMT optimal coding approaches are discussed: rotated lattices code \cite{Oggier2004}, and permutation code \cite{Tavildar2006}. From the simulation examples, the proposed closed-form formulas for outage probabilities are nearly identical with the simulation curves in both high and low SNR regimes. The results also show that for zero multiplexing gain (i.e., with a fixed target rate), the RB based coded $f$-matching scheme has $\unit[2]{dB}$ SNR gains compared to the interleaved or TDMA based allocation in IEEE 802.16 and 3GPP LTE-A. The SNR gain also becomes greater as the multiplexing gain increases. For the chunk based coded $f$-matching scheme, it achieves a much greater diversity gain than the localized allocation scheme. Therefore, the proposed framework cannot only be applied to analyze the performance of existing multi-carrier multi-access systems but also provide powerful tools for designing practical channel resources allocation schemes for future multi-carrier multi-access systems.

The rest of this paper is organized as follows. Section \ref{sec:system_model} presents the system model and precise problem formulation. The RBG based maximum $f$-matching framework and the requirements on coding schemes are presented in Section \ref{sec:coded_f_matching}. Section \ref{sec:oer_dmr_analysis} presents the closed-form formulas for outage probabilities, OER, and DMR of the proposed framework. The practical issues, such as the optimal parallel channel resources allocation algorithm and asymptotic optimal coding schemes, are studied in Section \ref{sec:practical_issues}. Section \ref{sec:simulation_results} presents the simulation results to verify our theoretical results. Finally, Section \ref{sec:conclusions} concludes this paper.

\section{System Model and Problem Formulation}\label{sec:system_model}

\subsection{Multi-Carrier Multi-Access Channel Model}\label{subsec:channel_model}

\begin{figure*}[t]
\centering
\begin{tikzpicture}[>=stealth]
	\draw[thick] (0,2.8) rectangle (1.6,3.6);
	\node at (0.8,3.2) {BS};
	\draw[thick] (1.6,3.2) -- (2,3.2) -- (2,3.6);
	\draw[thick] (2,3.6) -- (2.3,3.9) -- (1.7,3.9) -- (2,3.6);
	
	\draw[thick] (9,0) rectangle (7.4,0.8);
	\node at (8.2,0.4) {User $u_{N}$};
	\draw[thick] (7.4,0.4) -- (7,0.4) -- (7,0.8);
	\draw[thick] (7,0.8) -- (7.3,1.1) -- (6.7,1.1) -- (7,0.8);
	
	\fill
	(8.2,1.5) circle (1.5pt)
	(8.2,1.8) circle (1.5pt)
	(8.2,2.1) circle (1.5pt);
	
	\draw[thick] (9,2.8) rectangle (7.4,3.6);
	\node at (8.2,3.2) {User $u_{2}$};
	\draw[thick] (7.4,3.2) -- (7,3.2) -- (7,3.6);
	\draw[thick] (7,3.6) -- (7.3,3.9) -- (6.7,3.9) -- (7,3.6);
	
	\draw[thick] (9,5.1) rectangle (7.4,5.9);
	\node at (8.2,5.5) {User $u_{1}$};
	\draw[thick] (7.4,5.5) -- (7,5.5) -- (7,5.9);
	\draw[thick] (7,5.9) -- (7.3,6.2) -- (6.7,6.2) -- (7,5.9);
	
	\draw[<->,very thick] (2.4,3.8) -- (6.6,1.1);
	\draw[<->,very thick] (2.4,3.9) -- (6.6,3.9);
	\draw[<->,very thick] (2.4,4) -- (6.6,6.2);

	\filldraw[fill=black!70] (-7,3.9) rectangle (-5.75,4.15);
	\filldraw[fill=black!70] (-7,5.15) rectangle (-5.75,5.4);
	\filldraw[fill=black!70] (-5.75,4.9) rectangle (-4.5,5.15);
	\filldraw[fill=black!70] (-5.75,5.65) rectangle (-4.5,5.9);
	\filldraw[fill=black!70] (-4.5,4.15) rectangle (-3.25,4.65);
	\filldraw[fill=black!70] (-4.5,5.4) rectangle (-3.25,5.65);
	\filldraw[fill=black!70] (-3.25,5.4) rectangle (-2,5.65);

	\filldraw[fill=black!40] (-7,4.15) rectangle (-5.75,5.15);
	\filldraw[fill=black!40] (-5.75,5.4) rectangle (-4.5,5.65);
	\filldraw[fill=black!40] (-4.5,3.9) rectangle (-3.25,4.15);
	\filldraw[fill=black!40] (-3.25,5.65) rectangle (-2,5.9);

	\filldraw[fill=black!10] (-7,5.4) rectangle (-5.75,5.65);
	\filldraw[fill=black!10] (-5.75,5.15) rectangle (-4.5,5.4);
	\filldraw[fill=black!10] (-5.75,3.9) rectangle (-4.5,4.15);
	\filldraw[fill=black!10] (-5.75,4.4) rectangle (-4.5,4.65);
	\filldraw[fill=black!10] (-4.5,4.9) rectangle (-3.25,5.15);
	\filldraw[fill=black!10] (-3.25,4.4) rectangle (-2,4.9);
	\filldraw[fill=black!10] (-3.25,5.15) rectangle (-2,5.4);
	
	\foreach \x in {-7,-5.75,-4.5,-3.25}
		\foreach \y in {3.9,4.15,4.4,4.65,4.9,5.15,5.4,5.65}
		{
			\draw (\x,\y) +(0,0) rectangle ++(1.25,.25);
		}
	
	\draw[->,very thick] (-7,3.9) -- (-1.5,3.9) node[right] {Time};
	\draw[->,very thick] (-7,3.9) -- (-7,6.3) node[above] {Frequency};
	\node at (-4.5,3.6) {RB based Allocation Scheme};
	\node at (-4,6.3) {$1$ Subchannel $=1$ RB};

	\node at (-7.3,5.65) {$\mathcal{S}_{4}^{\mathrm{c}}\,\big\{$};
	\node at (-7.3,5.15) {$\mathcal{S}_{3}^{\mathrm{c}}\,\big\{$};
	\node at (-7.3,4.65) {$\mathcal{S}_{2}^{\mathrm{c}}\,\big\{$};
	\node at (-7.3,4.15) {$\mathcal{S}_{1}^{\mathrm{c}}\,\big\{$};

	\filldraw[fill=black!70] (-7,0.3) rectangle (-2,0.8);
	\filldraw[fill=black!40] (-7,0.8) rectangle (-2,1.3);
	\filldraw[fill=black!10] (-7,1.3) rectangle (-2,1.8);
			
	\foreach \x in {-7,-5.75,-4.5,-3.25}
		\foreach \y in {0.3,0.55,0.8,1.05,1.3,1.55,1.8,2.05}
		{
			\draw (\x,\y) +(0,0) rectangle ++(1.25,.25);
		}

	\draw[->,very thick] (-7,0.3) -- (-1.5,0.3) node[right] {Time};
	\draw[->,very thick] (-7,0.3) -- (-7,2.7) node[above] {Frequency};
	\node at (-4.5,0) {Chunk based Allocation Scheme};
	\node at (-4,2.7) {$1$ Subchannel $=8$ RBs};
	
	\node at (-7.3,2.05) {$\mathcal{S}_{4}^{\mathrm{c}}\,\big\{$};
	\node at (-7.3,1.55) {$\mathcal{S}_{3}^{\mathrm{c}}\,\big\{$};
	\node at (-7.3,1.05) {$\mathcal{S}_{2}^{\mathrm{c}}\,\big\{$};
	\node at (-7.3,0.55) {$\mathcal{S}_{1}^{\mathrm{c}}\,\big\{$};

	\filldraw[fill=black!70] (0,0.8) rectangle (1.25,1.05);
	\node at (1.9,0.925) {User $u_{1}$};
	\filldraw[fill=black!40] (0,0.3) rectangle (1.25,0.55);
	\node at (1.9,0.425) {User $u_{2}$};
	\filldraw[fill=black!10] (2.75,0.8) rectangle (4,1.05);
	\node at (4.7,0.925) {User $u_{N}$};
	\draw (2.75,0.3) rectangle (4,0.55);
	\node at (4.92,0.425) {Unallocated};
\end{tikzpicture}
\caption{A multi-carrier multi-access system with one BS and $M\geq2$ users over the frequency-selective Rayleigh slow fading channel. BS allocates the subchannels in the set $\mathcal{S}_{m}$ to the user $u_{m}$ according to $\unit[1]{bit}$ CSI feedback.}\label{fig:system_desc}
\end{figure*}
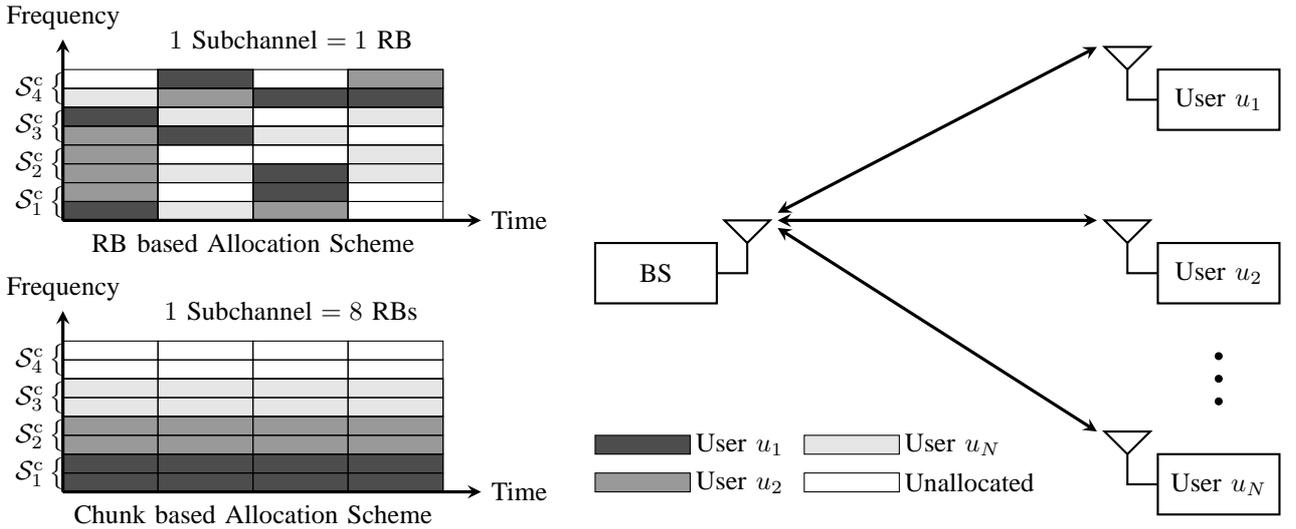

Consider a multi-carrier multi-access channel, as shown in Fig. \ref{fig:system_desc}, where the basic resource unit allocated to each user is referred to as the \emph{subchannel} in order to construct the unified analytic framework. Clearly, the subchannel is the aforementioned RB or chunk in the RB or chunk based allocation schemes, respectively.\footnote{The term ``RB'' or ``chunk'' will be used, if the term ``subchannel'' indicates one of them specifically.} In the considered multi-carrier multi-access channel, $M$ users communicate with a BS through $N$ subchannels with $N\geq M$. The user set is defined as $\mathcal{U}=\left\{u_{1},\ldots,u_{M}\right\}=\left\{u_{m}\right\}_{m=1}^{M}$, where $u_{m}$ denotes the $m$th user.\footnote{The script symbol $\mathcal{X}$ denotes a set, whose cardinality is denoted by $\left|\mathcal{X}\right|$.} The subchannel set in a given frame is defined as $\mathcal{S}=\left\{s_{n}\right\}_{n=1}^{N}$, where $s_{n}$ denotes the $n$th subchannel. The channel is assumed to be under-spread, i.e., the coherence time $T^{\mathrm{c}}$ is much larger than the multi-path delay spread $T^{\mathrm{d}}$ and also the length of each fame \cite{Proakis2007}. The signals of each user are assumed to undergo independent $L$-path frequency-selective fading, i.e., the channel contains $L$ coherence bandwidths. Let $\mathcal{S}_{l}^{\mathrm{c}},\,l=1,\ldots,L$ denote the set of subchannels in the $l$th coherence bandwidth, then $\left|\mathcal{S}_{l}^{\mathrm{c}}\right|=N_{\mathrm{c}}=\frac{N}{L}$, where the non-integer case is omitted since $N$ can be exactly divided by $L$ in the chunk based scheme or $N\gg L$ in the RB based scheme \cite{IEEE802.16-2009}. For convenience, the coherence bandwidth is used to normalize the total bandwidth, and the length of each frame is normalized as $1$.

According to the results in \cite{Wong1999}, it is quite reasonable to assume that the channel gains of the subchannels in one coherence bandwidth are the same, whereas they are independent with one another for different coherence bandwidths. Let $g_{ml}$ denote the channel gain of the $l$th coherence bandwidth for the user $u_{m}$. Then, $g_{ml}$, for every $m=1,\ldots,M$ and $l=1,\ldots,L$, are independent with the identical distribution of $\mathcal{CN}\left(0,1\right)$.\footnote{$\mathcal{CN}\left(\bm{\mu},\bm{C}\right)$ denotes a circularly symmetric Gaussian distribution with mean vector $\bm{\mu}$ and covariance matrix $\bm{C}$.} The channel gain on each subchannel, denoted by $h_{mn}$, can then be defined as $h_{mn}=g_{ml},\,\forall s_{n}\in\mathcal{S}_{l}^{\mathrm{c}}$. The mutual information of a subchannel $s_{n}\in\mathcal{S}_{l}^{\mathrm{c}}$ for the user $u_{m}$ is then given by
\begin{equation}\label{eq:subchannel_capacity}
	I_{mn}=\frac{1}{N_{\mathrm{c}}}\ln\left(1+\left|h_{mn}\right|^{2}\gamma\right)=\frac{1}{N_{\mathrm{c}}}\ln\left(1+\left|g_{ml}\right|^{2}\gamma\right),
\end{equation}
where $\gamma$ is the average SNR at the receiver. Throughout the paper, the natural logarithm function is used, and the unit of information is ``$\mathrm{nat}$''.

Because the subchannel must be allocated at the beginning of each frame in real-time, each user is only allowed to feedback $\unit[1]{bit}$ CSI for each coherence bandwidth so as to reduce the complexity in both signaling and subchannel allocation processes. Let $\bm{Q}$ denote the quantized CSI matrix at BS, where the element at the $m$th row and the $l$th column $\left[\bm{Q}\right]_{ml}$ is an $\unit[1]{bit}$ quantized value of $g_{ml}$. A subchannel allocation scheme, denoted by $\mathscr{S}$, can be seen as a mapping from $\bm{Q}$ to a family of subchannel allocation sets, that is
\begin{equation}\label{eq:subchannel_allocation}
	\left(\mathcal{S}_{m}\right)_{m=1}^{M}=\mathscr{S}\left(\bm{Q}\right),
\end{equation}
where $\left(\mathcal{S}_{m}\right)_{m=1}^{M}=\left(\mathcal{S}_{1},\ldots,\mathcal{S}_{M}\right)$ is a vector of sets, and $\mathcal{S}_{m}$ is the set of subchannels allocated to the user $u_{m}$, which satisfies $\mathcal{S}_{m}\cap\mathcal{S}_{m'}=\emptyset,\,\forall m'\neq m$, and $\mathcal{S}_{\mathrm{alloc}}=\bigcup_{u_{m}\in\mathcal{U}}\mathcal{S}_{m}\subseteq\mathcal{S}$. Over the time duration of one frame, the received signal of each symbol $\bm{y}\in\mathbb{C}^{N}$ in the frequency domain is given by
\begin{equation}\label{eq:channel_model}
	\bm{y}=\bm{H}\bm{x}+\bm{w},
\end{equation}
where $\bm{w}\in\mathbb{C}^{N}$ is the additive white Gaussian noise with the distribution of $\mathcal{CN}\left(\bm{0},\bm{I}\right)$. The channel gain $\bm{H}$ is a diagonal matrix and $\left[\bm{H}\right]_{nn}=h_{mn}$, if and only if $s_{n}\in\mathcal{S}_{m}$. The transmission symbol of all the users is denoted by the vector $\bm{x}=\left(x_{n}\right)_{n=1}^{N}$, where $\bm{x}_{m}=\left(x_{n}\right)_{s_{n}\in\mathcal{S}_{m}}$ is the symbol vector transmitted by the user $u_{m}$. Let $\mathscr{C}$ denote the coding scheme in the frequency domain, then $\bm{x}_{m}=\mathscr{C}\left(\bm{b}_{m}\right)$, where $\bm{b}_{m}$ is the uncoded-symbol vector for user $u_{m}$.

The objective of this paper is to propose a unified analytic framework for designing the  $\mathscr{S}$ and choosing $\mathscr{C}$ for subchannel allocation in multi-carrier multi-access systems. In the following, we will first assume $\mathscr{C}$ has been properly chosen and focus on designing the optimal $\mathscr{S}$. Then, the coding scheme, which can meet the requirement of the optimal $\mathscr{S}$, will be discussed.

\subsection{Outage Exponent Region \& Diversity-Multiplexing Region}
As defined in \cite{Ozarow1994}, if the instantaneous mutual information is smaller than a target transmission rate, this channel is in outage. The results in compound channels showed that the reliable communication can be achieved if the channel is not in outage \cite{Csiczar1981}. Therefore, the outage probability, which establishes the relationship between the reliability and efficiency, is a key performance metric in slow fading channels \cite{Zheng2003}.

According to the QoS requirement of the considered multi-carrier multi-access channel, each user has a specific target rate, which is denoted by $R_{m}$. A given vector of target rates $\bm{R}=\left(R_{m}\right)_{m=1}^{M}$ will be referred to as an \emph{operating point} of the system. If $\bm{Q}$ is already known at the BS, $\left(\mathcal{S}_{m}\right)_{m=1}^{M}$ can be generated by a given subchannel allocation scheme $\mathscr{S}$. The coding scheme $\mathscr{C}$ is assumed to be properly chosen such that the sum mutual information of the subchannels allocated to each user can be achieved. For a given operating point $\bm{R}$, therefore, the outage probability of a user $u_{m}$ is defined by
\begin{equation}\label{eq:outage_definition}
	p_{m}^{\mathrm{out}}\left(R_{m}\right)=\Pr\left\{\left.\sum_{s_{n}\in\mathcal{S}_{m}}I_{mn}<R_{m}\right|\mathcal{S}_{m}\right\}.
\end{equation}
Eq. \eqref{eq:outage_definition} implies that different users may have different outage probabilities. To evaluate the comprehensive performance of a subchannel allocation scheme, we define the \emph{outage exponent region} (OER), which specifies the set of outage exponents that are simultaneously achievable by the users for a fixed operating point. Similar to the error exponent region proposed in \cite{Weng2008}, the following definition describes this problem formally.

\begin{figure}[t]
\centering
\begin{tikzpicture}[>=stealth]
	\fill[domain=0:4,gray!50] plot (\x,{2*sqrt(1-pow(\x,2)/pow(4,2))}) -- (0,0) -- (0,2);
	\draw[domain=0:4,thick] plot (\x,{2*sqrt(1-pow(\x,2)/pow(4,2))});
	\draw[domain=0:4.5,thick,dashed] plot (\x,{2.5*sqrt(1-pow(\x,2)/pow(4.5,2))});
	\node at (1.8,0.9) {$\mathcal{R}_\mathrm{oer}\left(R_{1},R_{2};\gamma\right)$};
	\node at (3.5,2.7) {Boundary of $\mathcal{R}_\mathrm{dmr}\left(r_{1},r_{2}\right)$};
	\node at (1.3,2.1) {$\gamma\rightarrow\infty$};

	\draw[->] (3.5,2.5) -- (3,1.8634);
	\draw[->] (1.8,1.7861) -- (2.1,2.2111);

	\draw[->,very thick] (0,0) -- (5,0) node[right] {$E_{1}\left(R_{1}\right)$};
	\draw[->,very thick] (0,0) -- (0,3) node[above] {$E_{2}\left(R_{2}\right)$};
\end{tikzpicture}
\caption{The OER $\mathcal{R}_{\mathrm{oer}}\left(R_{1},R_{2};\gamma\right)$ and DMR $\mathcal{R}_{\mathrm{dmr}}\left(r_{1},r_{2}\right)$ for an operating point $\left(R_{1},R_{2}\right)$.}\label{fig:region_example}
\end{figure}
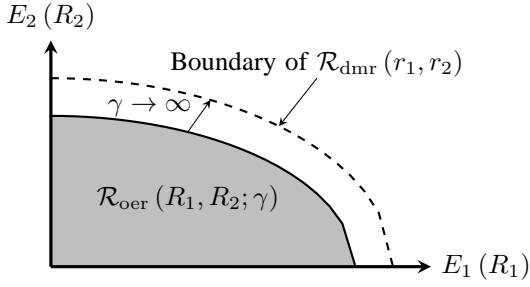

\medskip
\begin{defn}\label{def:OER}
For a given operating point $\bm{R}$, the OER, denoted by $\mathcal{R}_{\mathrm{oer}}\left(\bm{R};\gamma\right)$, consists of all vectors of outage exponents $\left(E_{m}\left(R_{m},\gamma\right)\right)_{m=1}^{M}$, which can be achieved by at least one subchannel allocation scheme $\mathscr{S}$. In this context, the outage exponent is defined as
\begin{equation}\label{eq:outage_exponent}
	E_{m}\left(R_{m},\gamma\right)=-\frac{\partial\ln p_{m}^{\mathrm{out}}\left(R_{m}\right)}{\partial\ln\gamma}.
\end{equation}
\end{defn}
\medskip

Fig. \ref{fig:region_example} shows an example of the OER for a subchannel allocation scheme operated at point $\left(R_{1},R_{2}\right)$, which is a two-dimensional region depending on the operating point. It should be noted that there is only one capacity region for a given system, but there is one OER for every operating point inside the capacity region. Clearly, a larger OER implies a better performance for a given operating point.

According to the results in \cite{Bai2011c}, the diversity-multiplexing tradeoff (DMT) for a user $u_{m}$ is closely related to the outage exponent as follows:
\begin{equation}\label{eq:diversity_gain}
	\lim_{\mathsf{\gamma\rightarrow\infty}}E_{m}\left(R_{m},\gamma\right)=d_{m}\left(r_{m}\right),
\end{equation}
where
\begin{equation}
	r_{m}=\lim_{\mathsf{\gamma\rightarrow\infty}}\frac{R_{m}}{\ln\left(1+\gamma\right)}
\end{equation}
is known as the multiplexing gain. As shown in Fig. \ref{fig:region_example}, the \emph{diversity-multiplexing region} (DMR), defined in the following, is the asymptotic case of the OER when $\gamma$ tends to infinity.

\medskip
\begin{defn}\label{def:DMR}
For a given operating point $\bm{r}=\left(r_{m}\right)_{m=1}^{M}$, the DMR, denoted by $\mathcal{R}_{\mathrm{dmr}}\left(\bm{r}\right)$, consists of all vectors of diversity gains $\left(d_{m}\left(r_{m}\right)\right)_{m=1}^{M}$, which can be achieved by at least one subchannel allocation scheme $\mathscr{S}$.
\end{defn}
\medskip

The optimal DMT for parallel fading channels is given by $d^{*}\left(r\right)=L\left(1-r/L\right)$ \cite{Bai2011c}. Therefore, for the DMR of the considered multi-carrier multi-access channel, the element of the optimal DMT vector should be given by
\begin{equation}\label{eq:opt_dmt}
	d_{m}^{*}\left(r_{m}\right)=L\left(1-\frac{r_{m}}{r_{m}^{*}}\right),\quad m=1,\ldots,M,
\end{equation}
where
\begin{equation}\label{eq:opt_dmt_cond}
\left\{\begin{aligned}
	& \frac{r_{1}^{*}}{R_{1}}=\cdots=\frac{r_{M}^{*}}{R_{M}},\\
	& \sum_{u_{m}\in\mathcal{U}}r_{m}^{*}=L.
\end{aligned}\right.
\end{equation}
In fact, Eqs. \eqref{eq:opt_dmt} \eqref{eq:opt_dmt_cond} indicate that each user shares the total multiplexing gain according to the operating point, while all of them achieve the full frequency diversity. Thus, a subchannel allocation scheme is referred to be \emph{asymptotic optimal}, denoted by $\mathscr{S}^{*}$, if and only if Eqs. \eqref{eq:opt_dmt} \eqref{eq:opt_dmt_cond} are achieved.

\section{RBG based Coded $f$-Matching Framework}\label{sec:coded_f_matching}

In this section, the maximum $f$-matching approach will first be proposed for designing the asymptotic optimal subchannel allocation schemes $\mathscr{S}^{*}$. The coding scheme $\mathscr{C}$ which fulfills the requirement of the proposed framework, denoted by $\mathscr{C}^{*}$, will also be discussed.

\subsection{Random Bipartite Graph Formulation}\label{subsec:RBG_Formulation}

Based on the aforementioned $\bm{Q}$ matrix, the considered multi-carrier multi-access channel can be formulated as a \emph{random bipartite graph} (RBG) model. In order to satisfy Eq. \eqref{eq:opt_dmt_cond}, it requires every subchannel to contribute the same amount of efficiency and reliability to each user. This requirement implies that all of the elements in $\bm{Q}$ should be independent and follows the identical distribution. The subchannel set $\mathcal{S}$ will then be shared by all of the users according to the operating point of the considered multi-carrier multi-access channel. Define
\begin{equation}\label{eq:subchannel_rate}
	R_{\mathrm{s}}=\frac{1}{N}\sum_{u_{m}\in\mathcal{U}}R_{m}.
\end{equation}
If $I_{mn}<R_{\mathrm{s}}$, we let $\left[\bm{Q}\right]_{ml}=0,\,\forall s_{n}\in\mathcal{S}_{l}^{\mathrm{c}}$, and refer this  subchannel as an outage subchannel for user $u_{m}$. Otherwise, $\left[\bm{Q}\right]_{ml}=1$. According to Eq. \eqref{eq:subchannel_capacity}, the subchannel outage probability is given by
\begin{equation}\label{eq:subchannel_outage}	p_{\mathrm{s}}=\Pr\left\{\left[\bm{Q}\right]_{ml}=0\right\}=1-\exp\left(-\frac{e^{N_{\mathrm{c}}R_{\mathrm{s}}}-1}{\gamma}\right).
\end{equation}
The non-outage probability of the subchannel is given by
\begin{equation}
	q_{\mathrm{s}}=1-p_{\mathrm{s}}.
\end{equation}
It can be seen that the subchannel outage probability is determined by $N_{\mathrm{c}}R_{\mathrm{s}}$, which is the target rate normalized by the coherence bandwidth. Since this parameter is very important in this paper, we let $R_{\mathrm{c}}=N_{\mathrm{c}}R_{\mathrm{s}}$ for convenience.

Note that $\bm{Q}$ denotes the independent outage state of each subchannel for each user at the target rate $R_{\mathrm{s}}$. In this context, each subchannel provides the same contribution of the rate and diversity to each user. Therefore, the user $u_{m}$ will be allocated
\begin{equation}
	\widetilde{K}_{m}=\frac{R_{m}}{R_{\mathrm{s}}}
\end{equation}
subchannels to guarantee the target rate. Recalling Eq. \eqref{eq:opt_dmt_cond}, we have $r_{m}^{*}=\frac{\widetilde{K}_{m}}{N_{\mathrm{c}}}$.

Before describing the RBG formulation in detail, we need to introduce some concepts in graph theory \cite{Bollobas1998}. A bipartite graph is a graph on which the set of vertices $\mathcal{V}$ admits a partition into two classes such that every edge has its ends in different class. A bipartite graph is often denoted by $\mathcal{G}\left(\mathcal{V}_{1}\cup\mathcal{V}_{2},\mathcal{E}\right)$, where $\mathcal{E}$ is the set of edges, $\mathcal{V}_{1}$ and $\mathcal{V}_{2}$ are the sets of vertices in different partition classes with $\mathcal{V}=\mathcal{V}_{1}\cup\mathcal{V}_{2}$ and $\mathcal{V}_{1}\cap\mathcal{V}_{2}=\emptyset$. A bipartite graph is called complete if every two vertices from different partition classes are adjacent. The complete bipartite graph with $M$ and $N$ vertices in different partition classes is denoted by $\mathcal{K}_{MN}$. Define a probability space $\left(\Omega,\mathscr{A},\mathsf{P}\right)$, in which the sample space contains all the bipartite graphs with $M$ vertices and $N$ vertices in different partition classes, $\mathscr{A}$ is the $\sigma$-field on $\Omega$, and $\mathsf{P}$ is the probability measure on $\mathscr{A}$. It is easy to verify that the sample space has $2^{MN}$ bipartite graphs. Similar to \cite{Bollobas2001}, the probability space of a RBG is denoted by $\mathscr{G}\left\{ \mathcal{K}_{MN};\mathsf{P}\right\}$ which means that every edge of the sample bipartite graph is chosen from $\mathcal{K}_{MN}$ with the probability measure $\mathsf{P}$. Clearly, a sample of $\mathscr{G}\left\{ \mathcal{K}_{MN};\mathsf{P}\right\}$ is a bipartite graph $\mathcal{G}\left(\mathcal{V}_{1}\cup\mathcal{V}_{2},\mathcal{E}\right)$. We also introduce some terms, which will be used in the following part. A vertex $v'$ is called a neighbor of vertex $v$, if they are adjacent. All the neighboring vertices of a given vertex $v$ are denoted by set $\mathcal{N}\left(v\right)\subset\mathcal{V}$, and the cardinality of $\mathcal{N}\left(v\right)$ is referred to as the degree of the vertex $v$, and denoted by $\deg_{\mathcal{G}}\left(v\right)$.

For the considered multi-carrier multi-access channel, the RBG model can be constructed as follows. First, all of the vertices in one partition class are used to represent all of the users in $\mathcal{U}$, and all of the vertices in the other partition class are used to represent all of the subchannels in $\mathcal{S}$, i.e., $\mathcal{V}_{1}=\mathcal{U}$ and $\mathcal{V}_{2}=\mathcal{S}$. If a subchannel $s_{n}\in\mathcal{S}$ is not in outage for a user $u_{m}\in\mathcal{U}$ according to the $\bm{Q}$ matrix, we join the vertices $u_{m}$ and $s_{n}$ with an edge $e_{mn}=u_{m}s_{n}$ in the RBG. Otherwise, if $s_{n}$ is in outage for $u_{m}$, there will be no edge between them. Thus, $\mathcal{N}\left(u_{m}\right)$ contains all of the subchannels that are not in outage for the user $u_{m}$. According to the construction method of $\bm{Q}$, the probability measure $\mathsf{P}$ can be defined as follows:
\begin{enumerate}
	\item Let $s_{n}\in\mathcal{S}_{l}^{\mathrm{c}}$ and $s_{n'}\in\mathcal{S}_{l'}^{\mathrm{c}}$, the edges $e_{mn}$ and $e_{m'n'}$ are chosen from $\mathcal{K}_{MN}$ independently with the identical probability $q_{\mathrm{s}}$, if and only if $\mathcal{S}_{l}^{\mathrm{c}}\neq\mathcal{S}_{l'}^{\mathrm{c}}$.
	\item The edges $e_{mn}$ and $e_{m'n'}$ are chosen from $\mathcal{K}_{MN}$ simultaneously, if and only if $\mathcal{S}_{l}^{\mathrm{c}}=\mathcal{S}_{l'}^{\mathrm{c}}$.
\end{enumerate}
Therefore, a sample of $\mathscr{G}\left\{\mathcal{K}_{MN};\mathsf{P}\right\}$ corresponds to a sample of $\bm{Q}$ and vice verse. A sample of $\mathscr{G}\left\{\mathcal{K}_{3,6};\mathsf{P}\right\}$ is shown in Fig. \ref{fig:RBG_example}.

\subsection{Maximum $f$-Matching based Subchannel Allocation}\label{sec:max_f_matching}

Given a sample of $\mathscr{G}\left\{\mathcal{K}_{MN};\mathsf{P}\right\}$, BS should allocate $\widetilde{K}_{m}$ subchannels to the user $u_{m}$ so as to guarantee Eqs. \eqref{eq:opt_dmt} \eqref{eq:opt_dmt_cond}. For designing the optimal subchannel allocation scheme, we introduce the concept of $f$-matching and $f$-factor in matching or factorization theory \cite{Lovasz1986}. Define a non-negative integer function $f\left(v\right)$ on vertex set $\mathcal{V}=\mathcal{U}\cup\mathcal{S}$, and let $w\left(e\right),\:\forall e\in\mathcal{E}$ denote an integral weight over edge $e$. Let $e_{v}$ denote the edge that is incident with the vertex $v$, the $\mathcal{M}_{f}$-degree of vertex $v$ is then given by
\begin{equation}
	\deg_{\mathcal{M}_{f}}\left(v\right)=\sum_{e_{v}\in\mathcal{M}_{f}}w\left(e_{v}\right).
\end{equation}
Therefore, the $f$-matching can be defined as follows.

\medskip
\begin{defn}\label{def:f_matching}
An $f$-matching $\mathcal{M}_{f}$ is a subset of $\mathcal{E}$ which satisfies
\begin{equation}\label{eq:f_matching}
	\deg_{\mathcal{M}_{f}}\left(v\right)\leq f\left(v\right),\quad\forall v\in\mathcal{V}.
\end{equation}
If $\deg_{\mathcal{M}_{f}}\left(v\right)=f\left(v\right)$, the vertex $v$ is referred to as the $\mathcal{M}_{f}$-saturated. The $f$-matching with the maximum number of edges is known as the maximum $f$-matching, denoted by $\mathcal{M}_{f}^{\mathrm{m}}$. Especially, the maximum $f$-matching is called the perfect $f$-matching or $f$-factor, denoted by $\mathcal{M}_{f}^{\mathrm{p}}$, if the equality of Eq. \eqref{eq:f_matching} holds for every vertex $v\in\mathcal{V}$.
\end{defn}
\medskip

Consider now the subchannel allocation problem in the considered multi-carrier multi-access channel. Let $w\left(e\right)=1,\,\forall e\in\mathcal{E}$, and define $f\left(v\right)$ as
\begin{equation}\label{eq:f-function}
	f\left(v\right)=
	\left\{\begin{aligned}
		& K_{m}, & v=u_{m}\in\mathcal{U};\\
		& 1, & v=s_{n}\in\mathcal{S}.
	\end{aligned}\right.
\end{equation}
In order to guarantee Eq. \eqref{eq:opt_dmt_cond}, the maximum $f$-matching $\mathcal{M}_{f}^{\mathrm{m}}$ should be generated to satisfy the following conditions:
\begin{equation}\label{eq:opt_cond_f-function}
	\frac{K_{1}}{\widetilde{K}_{1}}=\cdots=\frac{K_{M}}{\widetilde{K}_{M}}.
\end{equation}
For a generated maximum $f$-matching $\mathcal{M}_{f}^{\mathrm{m}}$, the number of non-outage subchannels allocated to the user $u_{m}$ is given by $k_{m}=\deg_{\mathcal{M}_{f}^{\mathrm{m}}}\left(u_{m}\right)$. Clearly, the following condition must also be guaranteed by the generated $\mathcal{M}_{f}^{\mathrm{m}}$:
\begin{equation}\label{eq:opt_cond_fmatching}
	\frac{k_{1}}{\widetilde{K}_{1}}=\cdots=\frac{k_{M}}{\widetilde{K}_{M}}.
\end{equation}
Here, we let $(\widetilde{K}_{\left(m\right)})_{m=1}^{M}$, $(K_{\left(m\right)})_{m=1}^{M}$, and $(k_{\left(m\right)})_{m=1}^{M}$ be the ordered vectors which are from the minimum value to the maximum one. For convenience, we also define
\begin{equation}\label{eq:K_sum}
\left\{\begin{aligned}
	& \widetilde{K}^{\mathrm{sum}}=\sum_{u_{m}\in\mathcal{U}}\widetilde{K}_{m}=N,\\
	& K^{\mathrm{sum}}=\sum_{u_{m}\in\mathcal{U}}K_{m},\\
	& k^{\mathrm{sum}}=\sum_{u_{m}\in\mathcal{U}}k_{m};
\end{aligned}\right.
\end{equation}
and
\begin{equation}
\left\{\begin{aligned}
	& \widetilde{K}_{m}^{\mathrm{sum}}=\widetilde{K}^{\mathrm{sum}}-\widetilde{K}_{m},\\
	& K_{m}^{\mathrm{sum}}=K^{\mathrm{sum}}-K_{m},\\
	& k_{m}^{\mathrm{sum}}=k^{\mathrm{sum}}-k_{m}.
\end{aligned}\right.
\end{equation}
If $k_{m}=K_{m}=\widetilde{K}_{m}$ for any $u_{m}\in\mathcal{U}$, all of the user are $\mathcal{M}_{f}$-saturated, i.e., no users are in outage. The maximum $f$-matching satisfying this condition will be referred to as the $\mathcal{U}$-perfect $f$-matching, denoted by $\mathcal{M}^{\mathcal{U}}_{f}$. Therefore, the optimal subchannel allocation scheme has two steps:
\begin{enumerate}
	\item BS generates the maximum $f$-matching $\mathcal{M}_{f}^{\mathrm{m}}$ that satisfies Eq. \eqref{eq:opt_cond_fmatching} on each sample bipartite graph of $\mathscr{G}\left\{ \mathcal{K}_{MN};\mathsf{P}\right\}$ so as to allocate $k_{m}$ non-outage subchannels to the user $u_{m}$.
	\item If $\mathcal{M}_{f}^{\mathrm{m}}\neq\mathcal{M}^{\mathcal{U}}_{f}$, BS randomly allocates $\widetilde{K}_{m}-k_{m}$ subchannels from the set of outage subchannels to the user $u_{m}\in\mathcal{U}$.
\end{enumerate}
According to Eq. \eqref{eq:f-function}, if a $\mathcal{U}$-perfect $f$-matching $\mathcal{M}^{\mathcal{U}}_{f}$ is generated in a bipartite graph, then the target rates of every user can be guaranteed. The $\mathcal{U}$-perfect $f$-matching, however, may not always be exist for every sample of $\mathscr{G}\left\{\mathcal{K}_{MN};\mathsf{P}\right\} $. For the case of $k_{m}<\widetilde{K}_{m}$, whether the user $u_{m}$ is in outage or not also depends on the applied coding scheme.

\begin{figure}[t]
	\centering
	\begin{tikzpicture}
		\fill
		(-2,5) circle (3pt)
		(-2,3) circle (3pt)
		(-2,1) circle (3pt);
		
		\node at (-2.5,5) {$u_{1}$};
		\node at (-2.5,3) {$u_{2}$};
		\node at (-2.5,1) {$u_{3}$};
		
		\draw[line width=0.5pt,style=dashed] (-2.8,0.5) rectangle (-1.7,5.5);
		\node at (-2.25,5.8) {Users};
				
		\fill
		(2,5.5) circle (3pt)
		(2,4.5) circle (3pt)
		(2,3.5) circle (3pt)
		(2,2.5) circle (3pt)
		(2,1.5) circle (3pt)
		(2,0.5) circle (3pt);		

		\node at (2.5,5.5) {$s_{1}$};
		\node at (2.5,4.5) {$s_{2}$};
		\node at (2.5,3.5) {$s_{3}$};
		\node at (2.5,2.5) {$s_{4}$};
		\node at (2.5,1.5) {$s_{5}$};
		\node at (2.5,0.5) {$s_{6}$};
		
		\draw[line width=0.5pt,style=dashed] (1.7,0) rectangle (2.8,6);
		\node at (2.25,6.3) {Subchannels};
		
		\draw [line width=2pt]
		(-2,5) -- (2,5.5)
		(-2,5) -- (2,2.5)
		(-2,3) -- (2,1.5)
		(-2,1) -- (2,4.5)
		(-2,1) -- (2,0.5);
		
		\draw [line width=0.8pt]
		(-2,5) -- (2,4.5)
		(-2,5) -- (2,3.5)
		(-2,5) -- (2,1.5)
		(-2,5) -- (2,0.5)
		(-2,3) -- (2,5.5)
		(-2,3) -- (2,4.5)
		(-2,3) -- (2,3.5)
		(-2,3) -- (2,2.5)
		(-2,3) -- (2,0.5)
		(-2,1) -- (2,5.5)
		(-2,1) -- (2,3.5)
		(-2,1) -- (2,2.5)
		(-2,1) -- (2,1.5);
	\end{tikzpicture}
	\caption{A sample of $\mathscr{G}\left\{\mathcal{K}_{3,6};\mathsf{P}\right\}$ with $K_{m}=\widetilde{K}_{m}=2,\,m=1,2,3$, where the outage probability of each edge is given by Eq. \eqref{eq:subchannel_outage}. The thick lines are the edges in the maximum $f$-matching $\mathcal{M}_{f}^{\mathrm{m}}$.}\label{fig:RBG_example}
\end{figure}
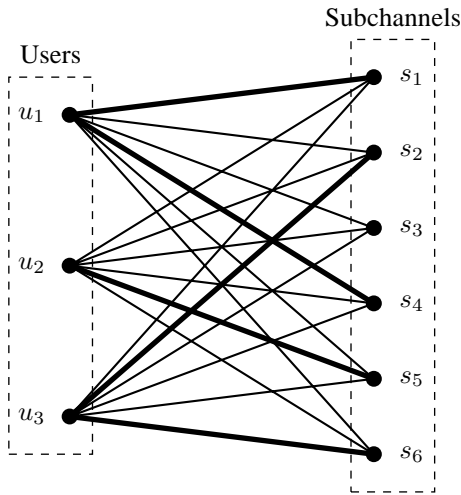

An example of the maximum $f$-matching based subchannel allocation is shown in Fig. \ref{fig:RBG_example}. The maximum $f$-matching $\mathcal{M}_{f}^{\mathrm{m}}=\left\{u_{1}s_{1},u_{1}s_{4},u_{2}s_{5},u_{3}s_{2},u_{3}s_{6}\right\}$ are denoted by thick lines. It can be seen that the users $u_{1}$ and $u_{3}$ are $\mathcal{M}_{f}^{\mathrm{m}}$-saturated, and thus are not in outage. Since the user $u_{2}$ is allocated one non-outage subchannel and one outage subchannel, its outage state will then be determined by the sum mutual information achieved by the applied coding scheme.

Clearly, the RBG based $f$-matching approach can be directly applied to both RB and chunk based allocation schemes, only if we see the RB or chunk as the subchannel.

\subsection{Requirements on Coding Schemes}\label{sec:coding}

According to the proposed maximum $f$-matching method, if the frequency-domain coding scheme $\mathscr{C}$ is not used, the outage state of a user is only determined by the relationship between $\widetilde{K}_{m}$ and $k_{m}$. In this context, the outage probability should be given by
\begin{equation}
	\Pr\left\{\left.k_{m}<\widetilde{K}_{m}\right|\mathcal{S}_{m}\right\},
\end{equation}
which is apparently greater than $p_{m}^{\mathrm{out}}$ defined in Eq. \eqref{eq:outage_definition}. Therefore, the maximum $f$-matching should be generated with the help of coding schemes to achieve the optimal outage performance.

It is well known that the minimum product distance between two codewords determines the capability of this coding scheme to combat the channel fading \cite{Tse2005}, which will determine the outage probability defined in Eq. \eqref{eq:outage_definition}. In the signal space of user $u_{m}$, each codeword can be seen as a vector or point in the $\widetilde{K}_{m}$-dimension space, i.e., $\widetilde{K}_{m}$-dimension constellation. If we choose any $s_{n}\in\mathcal{S}_{m}$, then the $n$th axis of the $\widetilde{K}_{m}$-dimension constellation corresponds to the channel gain $h_{mn}$. For convenience, let
\begin{equation}
\left\{\begin{aligned}
	& \bm{x}_{m}^{A}=\left(x_{mn}^{A}\right)_{s_{n}\in\mathcal{S}_{m}},\\
	& \bm{x}_{m}^{B}=\left(x_{mn}^{B}\right)_{s_{n}\in\mathcal{S}_{m}},
\end{aligned}\right.
\end{equation}
denote the codewords of the user $u_{m}$ for the information $A$ and $B$, respectively. The normalized product distance of $\bm{x}_{m}^{A}$ and $\bm{x}_{m}^{B}$ is then defined by
\begin{equation}\label{eq:product_distance}
	D_{m}=\frac{1}{\sqrt{\gamma}}\prod_{s_{n}\in\mathcal{S}_{m}}\left|x_{mn}^{A}-x_{mn}^{B}\right|.
\end{equation}

The coding scheme must be designed to maximize the minimum value of $D_{m}$, denoted by $D_{m}^{\mathrm{min}}$, so as to optimize the outage performance. If $D_{m}^{\mathrm{min}}=0$, for example $x_{n}^{A}=x_{n}^{B}$ for some $n$, there must be a hyperplane which is orthogonal with the $n$th axis such that $x_{n}^{A}-x_{n}^{B}$ is parallel with this hyperplane. Clearly, if all the channel gains on that hyperplane are very small, the receiver cannot distinguish $\bm{x}_{m}^{A}$ from $\bm{x}_{m}^{B}$. In other words, we lose one dimension, i.e., the $n$th subchannel, to combat the channel fading. Therefore, the number of terms that $x_{n}^{A}\neq x_{n}^{B},\,s_{n}\in\mathcal{S}_{m}$ in Eq. \eqref{eq:product_distance} is the diversity order achieved by this signal constellation \cite{Oggier2004}. This diversity order is another understanding, from the perspective of coding scheme, of the DMR in Definition \ref{def:DMR}. Therefore, the optimal coding scheme should be capable of achieving the optimal DMT in point-to-point parallel fading channels.

In this context, the coded $f$-matching based subchannel allocation scheme can be summarized as follows:
\begin{enumerate}
	\item Allocate subchannels to each user by using the maximum $f$-matching method in Section \ref{sec:max_f_matching};
	\item Apply the optimal coding scheme on the subchannel set $\mathcal{S}_{m}$, if $k_{m}<\widetilde{K}_{m}$ for the user $u_{m}$.
\end{enumerate}

\section{OER \& DMR Analysis for the Coded $f$-Matching Framework}\label{sec:oer_dmr_analysis}

\begin{table*}
  \centering
  \caption{Summarization of the Main Results}\label{tab:comparison}
  \begin{tabular}{|c|c|c|c|c|c|c|c|c|}
    \hline
    \multicolumn{2}{|c|}{Allocation Schemes} & OER & DMR & \begin{tabular}{c} Multi-User\\ Diversity \end{tabular} & \begin{tabular}{c} Allocation\\ Complexity \end{tabular} & \begin{tabular}{c} Coding\\ Length \end{tabular} & CSI & FOE\tabularnewline
    \hline
    \multirow{3}{*}{\begin{tabular}{c} RB\\ Based \end{tabular}} & Coded $f$-Matching & Eqs. \eqref{eq:outage_RB_high} \eqref{eq:outage_RB_LowSNR} & {\begin{tabular}{c} Eq. \eqref{eq:DMR_RB}\\ Optimal\end{tabular}} & Yes & $\mathcal{O}\left(\log^{2\eta}N\right)$ & $\frac{\widetilde{K}_{m}}{N_{\mathrm{c}}}$ & $\unit[1]{bit}$ & High\tabularnewline
    \cline{2-9}
    & Interleaved & Eq. \eqref{eq:outage_exponent_interleaved} & {\begin{tabular}{c} Eq. \eqref{eq:DMR_RB}\\ Optimal\end{tabular}} & No & $\mathcal{O}\left(1\right)$ & $L$ & --- & High\tabularnewline
    \hline
    \multirow{5}{*}{\begin{tabular}{c} Chunk\\ Based\end{tabular}} & Coded $f$-Matching & Eqs. \eqref{eq:outage_chunk_1} \eqref{eq:outage_RB_LowSNR} & {\begin{tabular}{c} Eq. \eqref{eq:DMR_chunk}\\ Suboptimal\end{tabular}} & Yes & $\mathcal{O}\left(\log^{2\eta}L\right)$ & $K_{m}$ & $\unit[1]{bit}$ & Middle\tabularnewline
    \cline{2-9}
    & Localized & Eq. \eqref{eq:outage_exponent_interleaved} & {\begin{tabular}{c} Eq. \eqref{eq:DMT_Lower}\\ Worst\end{tabular}} & No & $\mathcal{O}\left(1\right)$ & $K_{m}$ & --- & Low\tabularnewline
    \cline{2-9}
    & TDMA & Eq. \eqref{eq:outage_exponent_interleaved} & {\begin{tabular}{c} Eq. \eqref{eq:DMR_RB}\\ Optimal\end{tabular}} & No & $\mathcal{O}\left(1\right)$ & $L$ & --- & Middle\tabularnewline
    \hline
  \end{tabular}
\end{table*}

In this section, the OER and DMR are analyzed for the proposed coded $f$-matching framework. The tight upper bound of the conditional outage probability is first presented to illustrate the performance of the optimal coding schemes with $\unit[1]{bit}$ CSI feedback per coherence bandwidth. The properties of the maximum $f$-matching method are then studied in detail. With these results, the outage probability, OER and DMR of the proposed coded $f$-matching framework are presented for both RB and chunk based allocation schemes. The main results can be briefly summarized in Table \ref{tab:comparison}, where the complexity of frequency offset estimation (FOE) is also listed \cite{Beek1997} .

\subsection{Tight Upper Bound of the Conditional Outage Probability}

Let the optimal coding scheme $\mathcal{C}^{*}$ be used such that $D_{p}^{\mathrm{min}}$ has been achieved. Focus on the user $u_{m}$ with a generated maximum $f$-matching $\mathcal{M}_{f}^{\mathrm{m}}$, then the \emph{conditional outage probability} (COP) is given by
\begin{equation}\label{eq:con_outage_probability}
	p_{m}^{\mathrm{cop}}\left(R_{m}\left|\mathcal{D}_{K_{m}k_{m}}\right.\right)=\Pr\left\{\left.\sum_{n=1}^{K_{m}}I_{mn}<R_{m}\right|\mathcal{D}_{K_{m}k_{m}}\right\},
\end{equation}
where
\begin{equation}\label{eq:state_space}
\begin{aligned}
	\mathcal{D}_{K_{m}k_{m}}= & \left\{I_{m1},\ldots,I_{mK_{m}}\left|I_{m1}\geq R_{\mathrm{s}},\ldots,I_{mk_{m}}\geq R_{\mathrm{s}},\right.\right.\\
 	& \left.I_{m,k_{m}+1}<R_{\mathrm{s}},\ldots,I_{mK_{m}}<R_{\mathrm{s}}\right\}.
\end{aligned}
\end{equation}
According to the proposed coded $f$-matching framework, it is only needed to consider the case that different subchannel in $\mathcal{S}_{m}$ belongs to different coherence bandwidth. This condition is true because only the scheme of coding across coherence bandwidths is capable of providing diversity gains. Even under this condition, it is still not trivial to obtain the exact closed-form formula for Eq. \eqref{eq:con_outage_probability}. In this paper, the saddle-point approximation method, which has a good balance between accuracy and complexity \cite{Butler2007}, is applied to derive a tight upper bound of Eq. \eqref{eq:con_outage_probability}. Let $\beta_{m}=1-\frac{k_{m}}{K_{m}}$, and define a symbol ``$\lesssim$'' as
\begin{equation}\label{eq:symbol}
	f\left(x\right)\lesssim g\left(x\right)\Leftrightarrow
	\left\{\begin{aligned}
		& f\left(x\right)\leq g\left(x\right);\\
		& \lim_{x\rightarrow\infty}\frac{f\left(x\right)}{g\left(x\right)}=1.
	\end{aligned}\right.
\end{equation}
The tight upper bound of Eq. \eqref{eq:con_outage_probability} can then be summarized in the following theorem.

\medskip
\begin{thm}\label{thm:con_outage_exponent}
The exponentially tight upper bound of the conditional outage probability in Eq. \eqref{eq:con_outage_probability} is given by
\begin{equation}\label{eq:con_outage_exponent}
	\begin{aligned}
		& p_{m}^{\mathrm{cop}}\left(R_{m}\left|\mathcal{D}_{K_{m}k_{m}}\right.\right)\lesssim p_{m}^{\mathrm{upper}}\left(R_{m}\left|\mathcal{D}_{K_{m}k_{m}}\right.\right)\\
		& =\psi \exp\left(-K_{m}\left[\left(\ln\gamma-R_{\mathrm{c}}\right)J_{2}\left(\gamma\right)+J_{1}\left(\gamma\right)-J_{0}\left(\gamma\right)\right]\right),
	\end{aligned}
\end{equation}
where
\begin{equation}\label{eq:exponent_part1}
	\psi=\frac{1}{\sqrt{2\pi K_{m}\sigma^{2}}\lambda^{*}},
\end{equation}
\begin{equation}\label{eq:exponent_part2}
	\left\{\begin{aligned}
		& J_{2}\left(\gamma\right)=\lambda^{*};\\
		& J_{1}\left(\gamma\right)=\ln p_{\mathrm{s}}^{\beta_{m}}q_{\mathrm{s}}^{\left(1-\beta_{m}\right)};\\
		& J_{0}\left(\gamma\right)=\frac{1}{\gamma}+\left(1-\beta_{m}\right)\ln\Gamma\left(1-\lambda^{*},e^{R_{\mathrm{c}}}\gamma^{-1}\right)\\
		& +\beta_{m}\ln\left(\Gamma\left(1-\lambda^{*},\gamma^{-1}\right)-\Gamma\left(1-\lambda^{*},e^{R_{\mathrm{c}}}\gamma^{-1}\right)\right).
	\end{aligned}\right.
\end{equation}
In Eqs. \eqref{eq:exponent_part1} \eqref{eq:exponent_part2}, $\lambda^{*}$ and $\sigma^{2}$ satisfy Eq. \eqref{eq:exponent_equation} and Eq. \eqref{eq:exponent_parameter}, respectively. In these equations,
\begin{equation}
	\Gamma\left(z,\alpha\right)=\int_{\alpha}^{\infty}e^{-t}t^{z-1}dt
\end{equation}
is the incomplete Gamma function, and
\begin{equation}
\begin{aligned}
	& G_{p,q}^{m,n}\left(z\left|
	\begin{aligned}
		a_{1}, & \ldots,a_{p}\\
		b_{1}, & \ldots,b_{q}
	\end{aligned}\right.\right)=\\
	 & \frac{1}{2\pi i}\oint_{\mathcal{L}}\frac{\prod_{j=1}^{m}\Gamma\left(b_{j}-s\right)\prod_{j=1}^{n}\Gamma\left(1-a_{j}+s\right)}{\prod_{j=m+1}^{q}\Gamma\left(1-b_{j}+s\right)\prod_{j=n+1}^{p}\Gamma\left(a_{j}-s\right)}z^{s}ds,
\end{aligned}
\end{equation}
is the Meijer's $G$-function \cite{Gradshteyn2007}.
\end{thm}
\begin{IEEEproof}
See Appendix \ref{app:con_outage_exponent}.
\end{IEEEproof}
\medskip

\begin{figure*}[t]
{\footnotesize
\begin{equation}\label{eq:exponent_equation}
	\begin{aligned}
		R_{\mathrm{c}}= & \ln\gamma+(1-\beta_{m})\ln\frac{e^{R_{\mathrm{c}}}}{\gamma}+(1-\beta_{m})\frac{G_{2,3}^{3,0}\left(\frac{e^{R_{\mathrm{c}}}}{\gamma}\left|
		\begin{array}{ccc}
			1, & 1, & -\\
			0, & 0, & 1-\lambda^{*}
		\end{array}\right.\right)}{\Gamma\left(1-\lambda^{*},\frac{e^{R_{\mathrm{c}}}}{\gamma}\right)}\\
 		& +\beta_{m}\frac{\Gamma\left(1-\lambda^{*},\frac{1}{\gamma}\right)\ln\frac{1}{\gamma}-\Gamma\left(1-\lambda^{*},\frac{e^{R_{\mathrm{c}}}}{\gamma}\right)\ln\frac{e^{R_{\mathrm{c}}}}{\gamma}+G_{2,3}^{3,0}\left(\frac{1}{\gamma}\left|
	 	\begin{array}{ccc}
			1, & 1, & -\\
			0, & 0, & 1-\lambda^{*}
		\end{array}\right.\right)-G_{2,3}^{3,0}\left(\frac{e^{R_{\mathrm{c}}}}{\gamma}\left|
		\begin{array}{ccc}
			1, & 1, & -\\
			0, & 0, & 1-\lambda^{*}
		\end{array}\right.\right)}{\Gamma\left(1-\lambda^{*},\frac{1}{\gamma}\right)-\Gamma\left(1-\lambda^{*},\frac{e^{R_{\mathrm{c}}}}{\gamma}\right)}.
	\end{aligned}
\end{equation}}
\rule[0.5ex]{1\textwidth}{0.5pt}
\end{figure*}

\begin{figure*}[t]
{\footnotesize
\begin{equation}\label{eq:exponent_parameter}
	\begin{aligned}
		\sigma^{2}= & \beta_{m}\frac{\Gamma\left(1-\lambda^{*},\frac{1}{\gamma}\right)\left(\ln\frac{1}{\gamma}\right)^{2}-\Gamma\left(1-\lambda^{*},\frac{e^{R_{\mathrm{c}}}}{\gamma}\right)\left(\ln\frac{e^{R_{\mathrm{c}}}}{\gamma}\right)^{2}+2G_{2,3}^{3,0}\left(\frac{1}{\gamma}\left|
		\begin{array}{ccc}
			1, & 1, & -\\
			0, & 0, & 1-\lambda^{*}
		\end{array}\right.\right)\ln\frac{1}{\gamma}}{\Gamma\left(1-\lambda^{*},\frac{1}{\gamma}\right)-\Gamma\left(1-\lambda^{*},\frac{e^{R_{\mathrm{c}}}}{\gamma}\right)}\\
		& +\beta_{m}\frac{-2G_{2,3}^{3,0}\left(\frac{e^{R_{\mathrm{c}}}}{\gamma}\left|
		\begin{array}{ccc}
			1, & 1, & -\\
			0, & 0, & 1-\lambda^{*}
		\end{array}\right.\right)\ln\frac{e^{R_{\mathrm{c}}}}{\gamma}+2G_{3,4}^{4,0}\left(\frac{1}{\gamma}\left|
		\begin{array}{cccc}
			1, & 1, & 1, & -\\
			0, & 0, & 0, & 1-\lambda^{*}
		\end{array}\right.\right)-2G_{3,4}^{4,0}\left(\frac{e^{R_{\mathrm{c}}}}{\gamma}\left|
		\begin{array}{cccc}
			1, & 1, & 1, & -\\
			0, & 0, & 0, & 1-\lambda^{*}
		\end{array}\right.\right)}{\Gamma\left(1-\lambda^{*},\frac{1}{\gamma}\right)-\Gamma\left(1-\lambda^{*},\frac{e^{R_{\mathrm{c}}}}{\gamma}\right)}\\
		& -\beta_{m}\left(\frac{\Gamma\left(1-\lambda^{*},\frac{1}{\gamma}\right)\ln\frac{1}{\gamma}-\Gamma\left(1-\lambda^{*},\frac{e^{R_{\mathrm{c}}}}{\gamma}\right)\ln\frac{e^{R_{\mathrm{c}}}}{\gamma}+G_{2,3}^{3,0}\left(\frac{1}{\gamma}\left|
		\begin{array}{ccc}
			1, & 1, & -\\
			0, & 0, & 1-\lambda^{*}
		\end{array}\right.\right)-G_{2,3}^{3,0}\left(\frac{e^{R_{\mathrm{c}}}}{\gamma}\left|
		\begin{array}{ccc}
			1, & 1, & -\\
			0, & 0, & 1-\lambda^{*}
		\end{array}\right.\right)}{\Gamma\left(1-\lambda^{*},\frac{1}{\gamma}\right)-\Gamma\left(1-\lambda^{*},\frac{e^{R_{\mathrm{c}}}}{\gamma}\right)}\right)^{2}\\
		& +(1-\beta_{m})\frac{2G_{3,4}^{4,0}\left(\frac{e^{R_{\mathrm{c}}}}{\gamma}\left|
		\begin{array}{cccc}
			1, & 1, & 1, & -\\
			0, & 0, & 0, & 1-\lambda^{*}
		\end{array}\right.\right)}{\Gamma\left(1-\lambda^{*},\frac{e^{R_{\mathrm{c}}}}{\gamma}\right)}-(1-\beta_{m})\left(\frac{G_{2,3}^{3,0}\left(\frac{e^{R_{\mathrm{c}}}}{\gamma}\left|
		\begin{array}{ccc}
			1, & 1, & -\\
			0, & 0, & 1-\lambda^{*}
		\end{array}\right.\right)}{\Gamma\left(1-\lambda^{*},\frac{e^{R_{\mathrm{c}}}}{\gamma}\right)}\right)^{2}.
	\end{aligned}
\end{equation}}
\rule[0.5ex]{1\textwidth}{0.5pt}
\end{figure*}

Recalling the definition in Eq. \eqref{eq:diversity_gain}, the conditional DMT, which illustrates the contribution of the coding scheme in the asymptotic case, can be summarized as follows.

\medskip
\begin{cor}\label{cor:CDMT}
For the optimal coding scheme with $\unit[1]{bit}$ CSI feedback per coherence bandwidth, the conditional DMT for the user $u_{m}$ is given by
\begin{equation}\label{eq:dmt}
\begin{aligned}
	d_{m}^{\mathrm{con}}\left(r_{m}\left|\mathcal{D}_{K_{m}k_{m}}\right.\right) & =k_{m}\left(1-\frac{r_{m}}{K_{m}}\right)\\
	& =K_{m}\left(1-\beta_{m}\right)\left(1-\frac{r_{m}}{K_{m}}\right).
\end{aligned}
\end{equation}
\end{cor}
\medskip

\begin{rmk}
Corollary \ref{cor:CDMT} shows that the diversity gain will be decreased to $K_{m}\left(1-\beta_{m}\right)$, when the transmitter knows that $K_{m}\left(1-\beta_{m}\right)$ subchannels are not in outage. This phenomenon is quite counter-intuitive that the conventional point of view thinks CSI feedback will increase the communication performance. How to explain this? As a matter of fact, the observation of the channel in the transmitter will result in the state space collapse of the channel gains. If the transmitter does not observe the channel, we have no choice but evaluate the outage probability in the full state space, i.e., the prior probability space. When the transmitter knows CSI, the outage probability will be evaluated in a restricted state space, i.e., the posterior probability space. For the partial CSI feedback, the sate space will be collapsed to $\mathcal{D}_{K_{m}k_{m}}$ as defined in Eq. \eqref{eq:state_space}. With perfect CSI feedback, the state space will be reduced to one element, i.e., whether the channel is in outage or not is a deterministic event. Therefore, the conditional outage performance can be seen as a ``local'' performance metric in the time domain. Let $I^{t}_{mn}$ and $\mathcal{D}^{t}_{K_{m}k_{m}}$ denote the mutual information and CSI feedback at timeslot $t$, respectively. In the long run, the outage probability of the system, which is a ``global'' performance metric, can be given by
\begin{equation}
	p_{m}^{\mathrm{out}}\left(R_{m}\right)=\frac{1}{T}\sum_{t=1}^{T}\Pr\left\{\left.\sum_{n=1}^{K_{m}}I^{t}_{mn}<R_{m}\right|\mathcal{D}^{t}_{K_{m}k_{m}}\right\}.
\end{equation}
According to the law of large numbers, we have
\begin{equation}
\begin{aligned}
	& \begin{aligned}
	p_{m}^{\mathrm{out}}\left(R_{m}\right)= & \sum_{\mathcal{D}_{K_{m}k_{m}}\in\mathscr{D}_{m}}\Pr\left\{\mathcal{D}_{K_{m}k_{m}}\right\}\cdot\\
	& \Pr\left\{\left.\sum_{n=1}^{\widetilde{K}_{m}}I_{mn}<R_{m}\right|\mathcal{D}_{K_{m}k_{m}}\right\}
	\end{aligned}\\
	& = \sum_{\mathcal{D}_{K_{m}k_{m}}\in\mathscr{D}_{K_{m}k_{m}}}p_{m}^{\mathrm{cop}}\left(R_{m}|\mathcal{D}_{K_{m}k_{m}}\right)\Pr\left\{\mathcal{D}_{K_{m}k_{m}}\right\},
\end{aligned}
\end{equation}
where $\mathscr{D}_{K_{m}k_{m}}$ is the space constructed by all possible $\mathcal{D}_{K_{m}k_{m}}$. Since
\begin{equation}
	\Pr\left\{\mathcal{D}_{K_{m}k_{m}}\right\}=p_{\mathrm{s}}^{K_{m}\beta_{m}}q_{\mathrm{s}}^{K_{m}\left(1-\beta_{m}\right)},
\end{equation}
then in the high SNR regime the ``global'' outage probability will converge to
\begin{equation}
\begin{aligned}
	p_{m}^{\mathrm{out}}\left(R_{m}\right) & \approx\gamma^{-K_{m}\left(1-\beta_{m}\right)\left(1-\frac{r_{m}}{K_{m}}\right)}p_{\mathrm{s}}^{K_{m}\beta_{m}}\\
	& \approx\gamma^{K_{m}\left(1-\frac{r_{m}}{K_{m}}\right)},
\end{aligned}
\end{equation}
which is just the ``global'' DMT in \cite{Bai2011c}.
\end{rmk}

\subsection{Properties of the Maximum $f$-Matching}

Theorem \ref{thm:con_outage_exponent} and Corollary \ref{cor:CDMT} presents the outage performance achieved by the optimal coding scheme with $\unit[1]{bit}$ CSI feedback per coherence bandwidth. To analyze the outage performance of the proposed coded $f$-matching framework, we still need to study the properties of the maximum $f$-matching on the RBG formulation of the considered multi-carrier multi-access channel. As a powerful tool for studying these properties, the \emph{max-flow min-cut} theorem is listed as Lemma \ref{lem:max_flow_min_cut} in the following \cite{Bollobas1998}.
\medskip
\begin{lem}\label{lem:max_flow_min_cut}
For a directed graph with source $\theta$ and sink $\zeta$, the maximum flow value from $\theta$ to $\zeta$ is equal to the minimum of the capacities of cuts separating $\theta$ from $\zeta$.
\end{lem}
\medskip

Based on the max-flow min-cut theorem in Lemma \ref{lem:max_flow_min_cut}, the following lemma can be obtained.

\medskip
\begin{lem}\label{lem:f_matching}
Let $\mathcal{G}\left(\mathcal{U}\cup\mathcal{S},\mathcal{E}\right)$ be a bipartite graph with a non-negative integer function $f\left(v\right)$ defined on $\mathcal{U}\cup\mathcal{S}$. Then, the bipartite graph $\mathcal{G}\left(\mathcal{U}\cup\mathcal{S},\mathcal{E}\right)$ has a maximum $f$-matching $\mathcal{M}_{f}^{\mathrm{m}}$ which cannot saturate a given vertex $u_{m}$, if and only if there is a subset $\mathcal{X}$ of $\mathcal{U}$ with $u_{m}\in\mathcal{X}$ satisfying
\begin{equation}\label{eq:f_matching_cond}
	\sum_{u_{i}\in\mathcal{X}}f\left(u_{i}\right)>\sum_{s_{n}\in\mathcal{N}\left(\mathcal{X}\right)}f\left(s_{n}\right),
\end{equation}
where $\mathcal{N}\left(\mathcal{X}\right)\subseteq\mathcal{S}$ is the adjacent vertex set of $\mathcal{X}$.
\end{lem}
\begin{IEEEproof}
See Appendix \ref{app:f_matching}.
\end{IEEEproof}
\medskip

From this lemma, the sufficient and necessary conditions for the existence of a specific maximum $f$-matching can be summarized as follows, where $K_{m}^{\mathrm{th}}$ is an integer such that
\begin{equation}
	K_{m}^{\mathrm{th}}=N+1-\left\lceil\frac{M}{M-1}K_{m}^{\mathrm{sum}}\right\rceil.
\end{equation}

\medskip
\begin{lem}\label{lem:matching_edges}
Let $\mathcal{G}\left(\mathcal{U}\cup\mathcal{S},\mathcal{E}\right)$ be a bipartite graph with $\left|\mathcal{U}\right|=M$, $\left|\mathcal{S}\right|=N$ and $2\leq M\leq N$. Define a non-negative integer function $f\left(v\right)$ as Eq. \eqref{eq:f-function}, then for any edge set $\mathcal{E}$ if and only if
\begin{enumerate}
   \item $|\mathcal{E}|\geq\left(M-1\right)N+K_{m}$ for the case of $K_{m}=1,\ldots,K_{m}^{\mathrm{th}}$,
   \item or $|\mathcal{E}|\geq M\left(K^{\mathrm{sum}}-1\right)+1$ for the case of $K_{m}=K_{m}^{\mathrm{th}}+1,\ldots,\widetilde{K}_{m}$;
\end{enumerate}
there must be a maximum $f$-matching in $\mathcal{G}\left(\mathcal{U}\cup\mathcal{S},\mathcal{E}\right)$ which can saturate the vertices $(u_{\left(i\right)})_{i=1}^{m}$, where $u_{\left(i\right)}$ has the same order as $K_{\left(i\right)}$, and $K^{\mathrm{sum}}$ is defined in Eq. \eqref{eq:K_sum}.
\end{lem}
\begin{IEEEproof}
See Appendix \ref{app:matching_edges}.
\end{IEEEproof}
\medskip

\subsection{OER \& DMR for RB based Coded $f$-Matching Scheme}

As shown in Fig. \ref{fig:system_desc}, the subchannels, i.e., RBs themselves, are directly allocated to each user in the RB based coded $f$-matching allocation scheme. According to the channel model, the number of outage RBs is the integral multiples of the number of RBs in one coherence bandwidth, i.e., the integral multiples of $N_{\mathrm{c}}$. Moreover, each user will be allocated $\widetilde{K}_{m}$ RBs in this scheme so as to achieve the optimal performance. The exact performance analysis is very difficult for the coded $f$-matching approach, we will focus on the first order approximations in the high and low SNR regimes. Based on Theorem \ref{thm:con_outage_exponent} and Lemma \ref{lem:matching_edges}, the main results are summarized in the following theorem.

\medskip
\begin{thm}\label{thm:outage_RB}
For the considered multi-carrier multi-access channel with $M$ users ($M\geq2$), $L$ coherence bandwidths, and $N$ RBs, if the RB based coded $f$-matching allocation scheme is applied, the achieved outage probability for the user $u_{m}$, in the high SNR regime, is given by
\begin{equation}\label{eq:outage_RB_high}
\begin{aligned}
	p_{m}^{\mathrm{out}}\left(R_{m}\right)= & \sum_{\kappa=L-\left\lceil\frac{\widetilde{K}_{m}}{N_{\mathrm{c}}}\right\rceil+1}^{L}\frac{L!p_{m}^{\mathrm{cop}}\left(R_{m}\left|\mathcal{D}_{\widetilde{K}_{m},L-\kappa}\right.\right)p_{\mathrm{s}}^{\kappa}}{\kappa!\left(L-\kappa\right)!}\\
	& +\mathsf{O}\left(p_{\mathrm{s}}^{L}\right),
\end{aligned}
\end{equation}
\end{thm}
\begin{IEEEproof}
See Appendix \ref{app:outage_RB}.
\medskip

According to Definition \ref{def:DMR} and Corollary \ref{cor:CDMT}, the achieved DMR of the RB based coded $f$-matching allocation scheme is obtained from Theorem \ref{thm:outage_RB}.

\medskip
\begin{thm}\label{thm:DMR_RB}
For an operating point $\bm{r}=\left(r_{m}\right)_{m=1}^{M}$, the best achievable bound of the DMR $\mathcal{R}_{\mathrm{dmr}}\left(\bm{r}\right)$ is given by the vector $\left(d_{m}^{*}\left(r_{m}\right)\right)_{m=1}^{M}$ which satisfies
\begin{equation}\label{eq:DMR_RB}
	d_{m}^{*}\left(r_{m}\right)=L\left(1-\frac{N_{\mathrm{c}}r_{m}}{\widetilde{K}_{m}}\right),
\end{equation}
for $0<r_{m}\leq\widetilde{K}_{m}$.
\end{thm}
\medskip

\begin{figure}[t]
	\centering
	\includegraphics[width=3.6in]{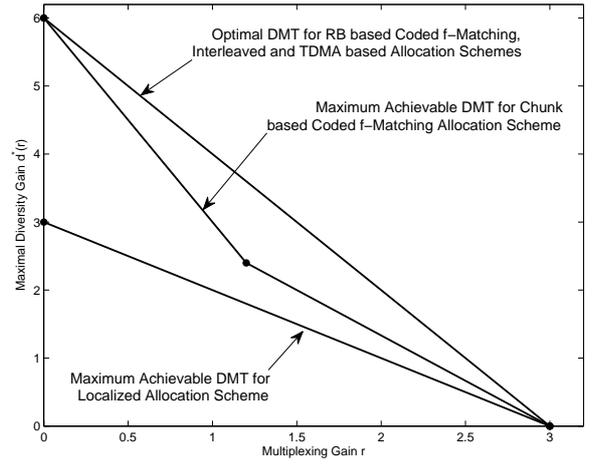}
	\caption{DMT curves of the user $u_{m}$ for different allocation schemes with $M=2$, $L=6$, and $N=12$.}\label{fig:DMTCurve}
\end{figure}

The DMT curve of the user $u_{m}$ and the DMR for the RB based coded $f$-matching allocation scheme are shown in Fig. \ref{fig:DMTCurve} and Fig. \ref{fig:OER}, respectively. Recalling Definition \ref{def:DMR} and Eqs. \eqref{eq:opt_dmt} \eqref{eq:opt_dmt_cond}, the RB based coded $f$-matching allocation scheme achieves the optimal DMR, such that all of the users share the multiplexing gain according to their target rates, and also achieve the full frequency diversity.

The previous results show the outage performance in the high SNR regime. A long distance between the user and the BS will lead to the low SNR regime. According to Eq. \eqref{eq:subchannel_outage}, we have $p_{\mathrm{s}}\rightarrow1$ as $\gamma\rightarrow0$, which indicates that almost all the RBs are in outage. Thus, the outage state caused by two users competing for one RB will nearly not happen. As a result, the outage state in the low SNR regime is mainly determined by the RB outage. This intuitive thinking is proved to be true by the following theorem.

\medskip
\begin{thm}\label{thm:outage_RB_LowSNR}
For the considered multi-carrier multi-access channel with $M$ users and $L$ coherence bandwidths, if the RB based coded $f$-matching allocation scheme is applied, the achieved outage probability for the user $u_{m}$, in the low SNR regime, is given by
\begin{equation}\label{eq:outage_RB_LowSNR}
	p_{m}^{\mathrm{out}}\left(R_{m}\right)=p_{\mathrm{s}}^{L}+Lp_{m}^{\mathrm{cop}}\left(R_{m}\left|\mathcal{D}_{K_{m},1}\right.\right)p_{\mathrm{s}}^{L-1}q_{\mathrm{s}}+\mathsf{O}\left(q_{\mathrm{s}}\right).
\end{equation}
\end{thm}
\begin{IEEEproof}
See Appendix \ref{app:outage_RB_LowSNR}.
\end{IEEEproof}
\medskip

Based on Theorem \ref{thm:outage_RB} and Theorem \ref{thm:outage_RB_LowSNR}, the OER of the RB based coded $f$-matching allocation scheme is summarized as the following theorem, which illustrate the comprehensive performance of the proposed scheme in the full SNR range.

\medskip
\begin{thm}\label{thm:OER_RB}
For an operating point $\bm{R}=\left(R_{m}\right)_{m=1}^{M}$, the best achievable bound of the OER $\mathcal{R}_{\mathrm{oer}}\left(\bm{R};\gamma\right)$ is given by the vector $\left(E_{m}\left(R_{m},\gamma\right)\right)_{m=1}^{M}$, which can be directly obtained by plugging Eq. \eqref{eq:outage_RB_high} and Eq. \eqref{eq:outage_RB_LowSNR} into Eq. \eqref{eq:outage_exponent}.
\end{thm}
\medskip

\begin{figure}[t]
	\centering
	\includegraphics[width=3.6in]{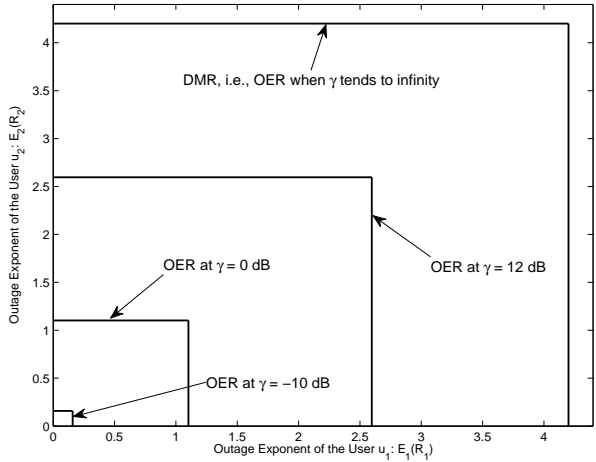}
	\caption{OER $\mathcal{R}_{\mathrm{oer}}\left(\bm{R};\gamma\right)$ and DMR $\mathcal{R}_{\mathrm{dmr}}\left(\bm{r}\right)$ for the RB based coded $f$-matching allocation scheme at operating point $\bm{r}=\left(0.9,0.9\right)$ with $M=2$, $L=6$, $N=12$.}\label{fig:OER}
\end{figure}

The DMR for the RB based coded $f$-matching allocation scheme at different SNRs are shown in Fig. \ref{fig:OER}. It can be seen that the OER approaches the DMR as SNR tends to infinity.

\begin{rmk}
There is a so called \emph{interleaved} allocation scheme in both IEEE 802.16 and LTE-A standards \cite{IEEE802.16-2009,3GPPLTE}. As shown in Fig. \ref{fig:interleaved}, one subcarrier in the duration of one frame is seen as one RB in this scheme, and the BS allocates RBs in different coherence bandwidths to users in an interleaved way.  As a matter of fact, the interleaved scheme is a special case of the RB based coded $f$-matching allocation scheme, whose OER and DMR performance can also be analysed by the code $f$-matching approach. For the user $u_{m}$, BS will allocate $\widetilde{K}_{m}$ RBs to it without observing the outage state of these RBs. According to the saddle-point approximation in the proof of Theorem \ref{thm:con_outage_exponent}, the exponentially tight upper bound of the outage probability $p_{m}^{\mathrm{out}}\left(R_{m}\right)$ is given by
\begin{equation}\label{eq:outage_exponent_interleaved}
	\begin{aligned}
		& p_{m}^{\mathrm{out}}\left(R_{m}\right)\lesssim p_{m}^{\mathrm{upper}}\left(R_{m}\right)\\
		& =\psi \exp\left(-\widetilde{K}_{m}\left[\left(\ln\gamma-R_{\mathrm{c}}\right)J_{2}\left(\gamma\right)-J_{0}\left(\gamma\right)\right]\right),
	\end{aligned}
\end{equation}
where
\begin{equation}
	\psi=\frac{1}{\sqrt{2\pi\widetilde{K}_{m}\sigma^{2}}\lambda^{*}},
\end{equation}
\begin{equation}
	\left\{\begin{aligned}
		& J_{2}\left(\gamma\right)=\lambda^{*};\\
		& J_{0}\left(\gamma\right)=\frac{1}{\gamma}+\ln\Gamma\left(1-\lambda^{*},\gamma^{-1}\right).
	\end{aligned}\right.
\end{equation}
The parameter $\lambda^{*}$ is the solution of the following equation:
\begin{equation}
	R_{\mathrm{c}}-\frac{1}{\Gamma\left(1-\lambda^{*},\gamma^{-1}\right)}G_{2,3}^{3,0}\left(\frac{1}{\gamma}\left|
	\begin{aligned} & 1,1\\
 			       & 0,0,1-\lambda^{*}
	\end{aligned}\right.\right)=0,
\end{equation}
while $\sigma^{2}$ is given by
\begin{equation}
	\begin{aligned}
		\sigma^{2}= & \frac{2}{\Gamma\left(1-\lambda^{*},\gamma^{-1}\right)}G_{3,4}^{4,0}\left(\frac{1}{\gamma}\left|\begin{aligned}
		     	& 1,1,1\\
			& 0,0,0,1-\lambda^{*}
		     \end{aligned}\right.\right)-\\
	 & \left[\frac{1}{\Gamma\left(1-\lambda^{*},\gamma^{-1}\right)}G_{2,3}^{3,0}\left(\frac{1}{\gamma}\left|\begin{aligned}
	& 1,1\\
	& 0,0,1-\lambda^{*}
\end{aligned}\right.\right)\right]^{2}.
	\end{aligned}
\end{equation}
It can be shown that the outage probability in Eq. \eqref{eq:outage_RB_high} is smaller than Eq, \eqref{eq:outage_exponent_interleaved}. Moreover, the achieved DMT of the interleaved allocation scheme is also shown to be given by Eq. \eqref{eq:DMR_RB}, if and only if $\widetilde{K}_{m}$ RBs are uniformly distributed in $L$ coherence bandwidths. The DMT curve of the user $u_{m}$ is plotted in Fig. \ref{fig:DMTCurve} for comparison.
\end{rmk}

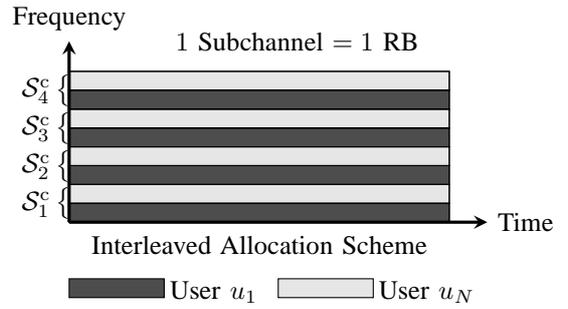
\begin{figure}[t]
\centering
\begin{tikzpicture}[>=stealth]
	\filldraw[fill=black!70] (-7,0.3) rectangle (-2,0.55);
	\filldraw[fill=black!10] (-7,0.55) rectangle (-2,0.8);
	\filldraw[fill=black!70] (-7,0.8) rectangle (-2,1.05);
	\filldraw[fill=black!10] (-7,1.05) rectangle (-2,1.3);
	\filldraw[fill=black!70] (-7,1.3) rectangle (-2,1.55);
	\filldraw[fill=black!10] (-7,1.55) rectangle (-2,1.8);
	\filldraw[fill=black!70] (-7,1.8) rectangle (-2,2.05);
	\filldraw[fill=black!10] (-7,2.05) rectangle (-2,2.3);
			
	\foreach \x in {-7}
		\foreach \y in {0.3,0.55,0.8,1.05,1.3,1.55,1.8,2.05}
		{
			\draw (\x,\y) +(0,0) rectangle ++(5,.25);
		}

	\draw[->,very thick] (-7,0.3) -- (-1.5,0.3) node[right] {Time};
	\draw[->,very thick] (-7,0.3) -- (-7,2.7) node[above] {Frequency};
	\node at (-4.5,0) {Interleaved Allocation Scheme};
	\node at (-4,2.7) {$1$ Subchannel $=1$ RB};
	
	\node at (-7.3,2.05) {$\mathcal{S}_{4}^{\mathrm{c}}\,\big\{$};
	\node at (-7.3,1.55) {$\mathcal{S}_{3}^{\mathrm{c}}\,\big\{$};
	\node at (-7.3,1.05) {$\mathcal{S}_{2}^{\mathrm{c}}\,\big\{$};
	\node at (-7.3,0.55) {$\mathcal{S}_{1}^{\mathrm{c}}\,\big\{$};
	
	\filldraw[fill=black!70] (-7,-0.7) rectangle (-5.75,-0.45);
	\node at (-5.1,-0.625) {User $u_{1}$};
	\filldraw[fill=black!10] (-4.25,-0.7) rectangle (-3,-0.45);
	\node at (-2.3,-0.625) {User $u_{N}$};
\end{tikzpicture}
\caption{The interleaved allocation scheme in IEEE 802.16 and LTE-A, where the RBs in different coherence bandwidths are allocated to different users in an interleaved way.}\label{fig:interleaved}
\end{figure}

\subsection{OER \& DMR for Chunk based Coded $f$-Matching Scheme}

The chunk based coded $f$-matching allocation scheme is considered in this subsection, where we assume $L>M$. The case of $L\leq M$ has been shown to achieve the optimal OER and DMR without requiring the frequency-domain coding scheme in \cite{Bai2011b}. As shown in Fig. \ref{fig:system_desc}, all of the RBs in each coherence bandwidth will be bundled as one chunk, i.e., $N=L$, because:  1) dividing each coherence bandwidth into more chunks can be seen as a special case of RB based coded $f$-matching allocation scheme; 2) bundling all of the RBs in multiple coherence bandwidths as one chunk must be worse than allocating more chunks in the proposed bundling method. Due to the difficulty in calculating the exact formula of the outage probability, we still focus on the first order approximations in the high and low SNR regimes. Based on Theorem \ref{thm:con_outage_exponent} and Lemma \ref{lem:matching_edges}, the main results are summarized in the following theorem.

\medskip
\begin{thm}\label{thm:outage_chunk}
For the considered multi-carrier multi-access channel with $M$ users and $L$ coherence bandwidths ($M\leq L$), if the chunk based coded $f$-matching allocation scheme is applied, the achieved outage probability for the user $u_{m}$, in the high SNR regime, is given by
\begin{equation}\label{eq:outage_chunk_1}
\begin{aligned}
	p_{m}^{\mathrm{out}}\left(R_{m}\right)= & \sum_{\kappa=L-K_{m}+1}^{L}\frac{L!p_{m}^{\mathrm{cop}}\left(R_{m}\left|\mathcal{D}_{K_{m},L-\kappa}\right.\right)p_{\mathrm{s}}^{\kappa}}{\kappa!\left(L-\kappa\right)!}\\
	& +\mathsf{O}\left(p_{\mathrm{s}}^{L}\right),
\end{aligned}
\end{equation}
for the case of $K_{m}=1,\ldots,K_{m}^{\mathrm{th}}$; or
\begin{equation}\label{eq:outage_chunk_2}
\begin{aligned}
   p_{m}^{\mathrm{out}}\left(R_{m}\right)= & \frac{L!p_{m}^{\mathrm{cop}}\left(R_{m}\left|\mathcal{D}_{K_{m},K_{m}-1}\right.\right)p_{\mathrm{s}}^{M\left(L-K^{\mathrm{sum}}+1\right)}}{M\left(L-K^{\mathrm{sum}}+1\right)!\left(K^{\mathrm{sum}}-1\right)!}\\
   & +\mathsf{O}\left(p_{\mathrm{s}}^{M\left(L-K^{\mathrm{sum}}+1\right)+K_{m}-1}\right),
\end{aligned}
\end{equation}
for the case of $K_{m}=K_{m}^{\mathrm{th}}+1,\ldots,\widetilde{K}_{m}$. For the case of $K_{m}=K_{m}^{\mathrm{th}}$ and $\left(M-1\right)|MK_{m}^{\mathrm{sum}}$, the outage probability of the user $u_{m}$ is the summation of Eq. \eqref{eq:outage_chunk_1} and Eq. \eqref{eq:outage_chunk_2}.
\end{thm}
\begin{IEEEproof}
See Appendix \ref{app:outage_chunk}.
\end{IEEEproof}
\medskip

Similar to Theorem \ref{thm:DMR_RB}, it is not difficult to obtain the following corollary from Theorem \ref{thm:outage_chunk} and Corollary \ref{cor:CDMT}.

\medskip
\begin{cor}\label{cor:DMT_chunk}
For the operating point $\left(r_{m}\right)_{m=1}^{M}$, the achieved DMT for the user $u_{m}$ is given by
\begin{equation}\label{eq:DMT_Lower}
	d_{m}\left(r_{m}\right)=L\left(1-\frac{r_{m}}{K_{m}}\right),
\end{equation}
for the case of $K_{m}=1,\ldots,K_{m}^{\mathrm{th}}$; and
\begin{equation}\label{eq:DMT_Upper}
	d_{m}\left(r_{m}\right)=\left[M\left(L-K^{\mathrm{sum}}+1\right)+K_{m}-1\right]\left(1-\frac{r_{m}}{K_{m}}\right),
\end{equation}
for the case of $K_{m}=K_{m}^{\mathrm{th}}+1,\ldots,\widetilde{K}_{m}$.
\end{cor}
\medskip

Both Theorem \ref{thm:outage_chunk} and Corollary \ref{cor:DMT_chunk} show that the chunk based coded $f$-matching allocation will not be optimal in OER and DMR. For a given user $u_{m}$, the best achievable DMT is a piece-wised linear function. In Eq. \eqref{eq:DMT_Lower}, the slope $-\frac{L}{K_{m}}$ increases as $K_{m}$ increasing from $1$ to $K_{m}^{\mathrm{th}}$, therefore
\begin{equation}\label{eq:OptDMT_Lower}
	d_{m}^{*}\left(r_{m}\right)=L\left(1-\frac{r_{m}}{K_{m}^{\mathrm{th}}}\right)
\end{equation}
outperforms the cases of $K_{m}=1,\ldots,K_{m}^{\mathrm{th}}-1$. In Eq. \eqref{eq:DMT_Upper}, similarly, the slope
\begin{equation}
\begin{aligned}
	& -\frac{M\left(L-K^{\mathrm{sum}}+1\right)+K_{m}-1}{K_{m}}\\
	= & -\frac{M\left(L-K_{m}^{\mathrm{sum}}+1\right)-1}{K_{m}}+M-1
\end{aligned}
\end{equation}
increases as $K_{m}$ increasing from $K_{m}^{\mathrm{th}}+1$ to $\widetilde{K}_{m}$. Thus,
\begin{equation}\label{eq:OptDMT_Upper}
	d_{m}^{*}\left(r_{m}\right)=\left[M\left(L-K^{\mathrm{sum}}+1\right)+\widetilde{K}_{m}-1\right]\left(1-\frac{r_{m}}{\widetilde{K}_{m}}\right)
\end{equation}
outperforms the cases of $K_{m}=K_{m}^{\mathrm{th}}+1,\ldots,\widetilde{K}_{m}$. Since $L>M\left(L-K^{\mathrm{sum}}+1\right)+\widetilde{K}_{m}-1$, a threshold of $r_{m}$ can be obtained by solving the following equation
\begin{equation}
\begin{aligned}
	& L\left(1-\frac{r_{m}}{K_{m}^{\mathrm{th}}}\right)\\
	= & \left[M\left(L-K^{\mathrm{sum}}+1\right)+\widetilde{K}_{m}-1\right]\left(1-\frac{r_{m}}{\widetilde{K}_{m}}\right).
\end{aligned}
\end{equation}
Clearly, we have
\begin{equation}
	r_{m}^{\mathrm{th}}=\frac{K_{m}^{\mathrm{th}}\widetilde{K}_{m}\left[L-\widetilde{K}_{m}+1-M\left(L-K^{\mathrm{sum}}+1\right)\right]}{\widetilde{K}_{m}\left(L-K_{m}^{\mathrm{th}}\right)+K_{m}^{\mathrm{th}}\left[1-M\left(L-K^{\mathrm{sum}}+1\right)\right]}.
\end{equation}
Therefore, the optimal chunk based coded $f$-matching allocation scheme within the specific bundle method can be described as follows. If the operating point for the user $u_{m}$ satisfies $r_{m}\leq r_{m}^{\mathrm{th}}$, the proposed coded $f$-matching method will allocate $K_{m}^{\mathrm{th}}$ chunks to this user; whereas if $r_{m}> r_{m}^{\mathrm{th}}$, the proposed method will allocate $\widetilde{K}_{m}$ chunks to this user. The best achievable DMT of the user $u_{m}$ is the piece-wised linear function joined by Eq. \eqref{eq:OptDMT_Lower} and Eq. \eqref{eq:OptDMT_Upper}. This result is summarized as follows.

\medskip
\begin{thm}\label{thm:DMR_chunk}
For an operating point $\bm{r}=\left(r_{m}\right)_{m=1}^{M}$, the best achievable bound of the DMR $\mathcal{R}_{\mathrm{dmr}}\left(\bm{r}\right)$ is given by the vector $\left(d_{m}^{*}\left(r_{m}\right)\right)_{m=1}^{M}$ which satisfies
\begin{equation}\label{eq:DMR_chunk}
	d_{m}^{*}\left(r_{m}\right)=L\left(1-\frac{r_{m}}{K_{m}^{\mathrm{th}}}\right),
\end{equation}
for $0<r_{m}\leq r_{m}^{\mathrm{th}}$; or
\begin{equation}
	d_{m}^{*}\left(r_{m}\right)=\left[M\left(L-K^{\mathrm{sum}}+1\right)+\widetilde{K}_{m}-1\right]\left(1-\frac{r_{m}}{\widetilde{K}_{m}}\right),
\end{equation}
for $r_{m}^{\mathrm{th}}<r_{m}<\widetilde{K}_{m}$.
\end{thm}
\medskip

The DMT curve of the user $u_{m}$ and the DMR for the chunk based coded $f$-matching allocation scheme are shown in Fig. \ref{fig:DMTCurve} and Fig. \ref{fig:OER}, respectively. It can be seen that the maximum achievable DMT for this scheme is not optimal. However, this scheme provides the multi-user diversity, and enjoys the low complexity of resource allocation algorithm and low requirement on frequency shift estimation.

The previous results show the outage performance in the high SNR regime. Consider now the low SNR regime, where almost all of the chunks are in outage. Similar to the RB based coded $f$-matching allocation scheme, the outage state in the low SNR regime is mainly determined by the chunk outage. Therefore, Theorem \ref{thm:outage_RB_LowSNR} can still be established for the chunk based coded $f$-matching allocation scheme.

According to Theorem \ref{thm:con_outage_exponent}, $p_{m}^{\mathrm{cop}}\left(R_{m}\left|\mathcal{D}_{K_{m}k_{m}}\right.\right)$ is a decreasing function of $K_{m}$. Therefore, the aforementioned analysis for the best achievable DMR is also applicable for OER. Let $\mathbf{1}_{\mathcal{X}}$ denote the indicative function of the event $\mathcal{X}$, then the result can be summarized as follows.

\medskip
\begin{thm}\label{thm:OER_chunk}
For an operating point $\bm{R}=\left(R_{m}\right)_{m=1}^{M}$, the best achievable bound of the OER $\mathcal{R}_{\mathrm{oer}}\left(\bm{R};\gamma\right)$ is given by the vector $\left(E_{m},\gamma\left(R_{m}\right)\right)_{m=1}^{M}$, which can be directly obtained by plugging
\begin{equation}\label{eq:Opt_outage_chunk_1}
\begin{aligned}
	& p_{m}^{\mathrm{out}}\left(R_{m}\right)=\sum_{\kappa=L-K_{m}^{\mathrm{th}}+1}^{L}\frac{L!p_{m}^{\mathrm{cop}}\left(R_{m}\left|\mathcal{D}_{K_{m},L-\kappa}\right.\right)p_{\mathrm{s}}^{\kappa}}{\kappa!\left(L-\kappa\right)!}\\
	& +\mathbf{1}_{\left\{\left(M-1\right)\left|MK_{m}^{\mathrm{th}}\right.\right\}}\frac{L!}{M\left(L-K^{\mathrm{sum}}+1\right)!\left(K^{\mathrm{sum}}-1\right)!}\cdot\\
	& p_{m}^{\mathrm{cop}}\left(R_{m}\left|\mathcal{D}_{K_{m},K_{m}^{\mathrm{th}}-1}\right.\right)p_{\mathrm{s}}^{M\left(L-K^{\mathrm{sum}}+1\right)}+\mathsf{O}\left(p_{\mathrm{s}}^{L}\right),
\end{aligned}
\end{equation}
into Eq. \eqref{eq:outage_exponent} for $0<r_{m}\leq r_{m}^{\mathrm{th}}$ in the high SNR regime; or
\begin{equation}\label{eq:Opt_outage_chunk_2}
\begin{aligned}
   p_{m}^{\mathrm{out}}\left(R_{m}\right)= & \frac{L!p_{m}^{\mathrm{cop}}\left(R_{m}\left|\mathcal{D}_{K_{m},\widetilde{K}_{m}-1}\right.\right)p_{\mathrm{s}}^{M\left(L-K^{\mathrm{sum}}+1\right)}}{M\left(L-K^{\mathrm{sum}}+1\right)!\left(K^{\mathrm{sum}}-1\right)!}\\
   & +\mathsf{O}\left(p_{\mathrm{s}}^{M\left(L-K^{\mathrm{sum}}+1\right)+\widetilde{K}_{m}-1}\right),
\end{aligned}
\end{equation}
into Eq. \eqref{eq:outage_exponent} for $r_{m}^{\mathrm{th}}<r_{m}<\widetilde{K}_{m}$ in the high SNR regime. In the low SNR regime, the OER can be obtained by plugging Eq. \eqref{eq:outage_RB_LowSNR} into Eq. \eqref{eq:outage_exponent}.
\end{thm}
\medskip

The DMR for the chunk based coded $f$-matching allocation scheme at different SNRs are similar to the regions in Fig. \ref{fig:OER}. Also, the OER approaches the DMR as SNR tends to infinity.

\begin{rmk}
There is another allocation scheme in both IEEE 802.16 and LTE-A standards, which is referred to as the \emph{localized} scheme \cite{IEEE802.16-2009,3GPPLTE}. As a matter of fact, the localized scheme is a fixed version of the chunk based coded $f$-matching allocation scheme in this subsection. Thus, the OER and DMR performance can also be analyzed by the proposed code $f$-matching approach. For the user $u_{m}$, BS will allocate $\widetilde{K}_{m}$ chunks to it without observing the outage state of these chunks. The outage probability $p_{m}^{\mathrm{out}}\left(R_{m}\right)$ is then exponentially tight upper bounded by Eq. \eqref{eq:outage_exponent_interleaved} with $\widetilde{K}_{m}$ being replaced by $\frac{\widetilde{K}_{m}}{N_{\mathrm{c}}}$ in the equation. The DMT curve of the user $u_{m}$ for the localized allocation scheme is plotted in Fig. \ref{fig:DMTCurve}. Compared to other allocation schemes, the localized allocation scheme can only achieve the worst DMT performance.
\end{rmk}

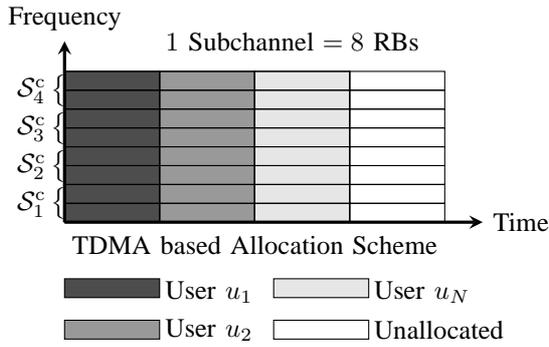
\begin{figure}[t]
\centering
\begin{tikzpicture}[>=stealth]
	\filldraw[fill=black!70] (-7,0.3) rectangle (-5.75,2.3);
	\filldraw[fill=black!40] (-5.75,0.3) rectangle (-4.5,2.3);
	\filldraw[fill=black!10] (-4.5,0.3) rectangle (-3.25,2.3);
			
	\foreach \x in {-7,-5.75,-4.5,-3.25}
		\foreach \y in {0.3,0.55,0.8,1.05,1.3,1.55,1.8,2.05}
		{
			\draw (\x,\y) +(0,0) rectangle ++(1.25,.25);
		}

	\draw[->,very thick] (-7,0.3) -- (-1.5,0.3) node[right] {Time};
	\draw[->,very thick] (-7,0.3) -- (-7,2.7) node[above] {Frequency};
	\node at (-4.5,0) {TDMA based Allocation Scheme};
	\node at (-4,2.7) {$1$ Subchannel $=8$ RBs};
	
	\node at (-7.3,2.05) {$\mathcal{S}_{4}^{\mathrm{c}}\,\big\{$};
	\node at (-7.3,1.55) {$\mathcal{S}_{3}^{\mathrm{c}}\,\big\{$};
	\node at (-7.3,1.05) {$\mathcal{S}_{2}^{\mathrm{c}}\,\big\{$};
	\node at (-7.3,0.55) {$\mathcal{S}_{1}^{\mathrm{c}}\,\big\{$};
	
	\filldraw[fill=black!70] (-7,-0.7) rectangle (-5.75,-0.45);
	\node at (-5.1,-0.625) {User $u_{1}$};
	\filldraw[fill=black!40] (-7,-1.25) rectangle (-5.75,-1);
	\node at (-5.1,-1.125) {User $u_{2}$};
	\filldraw[fill=black!10] (-4.25,-0.7) rectangle (-3,-0.45);
	\node at (-2.3,-0.625) {User $u_{N}$};
	\draw (-4.25,-1.25) rectangle (-3,-1);
	\node at (-2.08,-1.125) {Unallocated};
\end{tikzpicture}
\caption{TDMA is also a special case of the chunk based coded $f$-matching allocation scheme, where all of the RBs in one timeslot are bundled as one chunk.}\label{fig:TDMA}
\end{figure}

\begin{rmk}
Besides the bundling method shown in Fig. \ref{fig:system_desc}, another method, often referred to as TDMA, is shown in Fig. \ref{fig:TDMA}, where the RBs in one timeslot are bundled as one chunk. The OER and DMR performance can also be analyzed by the coded $f$-matching approach. For the user $u_{m}$, BS will allocate $\widetilde{K}_{m}$ chunks to it without observing the outage state of these chunks. Since the frame is in one coherence time, the outage probabilities are the same for each user and is exponentially tight upper bounded by Eq. \eqref{eq:outage_exponent_interleaved} with $\widetilde{K}_{m}$ being replaced by $L$ in the equation. The DMT curve of the user $u_{m}$ for TDMA scheme is plotted in Fig. \ref{fig:DMTCurve}. It can be seen that the optimal DMT can also be achieved by this scheme. However, this scheme has a high requirement on frequency shift estimation, and no multi-user diversity.
\end{rmk}

\section{Practical Issues}\label{sec:practical_issues}

The complexity of the proposed framework is a key issue for practical multi-carrier multi-access systems. In this section, the parallel maximum $f$-matching algorithm for subchannel allocation with log-polynomial complexity is discussed. The asymptotic optimal coding scheme, which is easy to implement, is also presented.

\subsection{Parallel Subchannel Allocation Algorithm}

The Hopcroft-Karp algorithm is well known for solving the maximum matching problem in bipartite graphs \cite{Hopcroft1973}. According to the matching theory \cite{Lovasz1986}, a matching $\mathcal{M}$ is maximum if and only if there is no augmenting path with respect to this matching. Consider a bipartite graph $\mathcal{G}\left(\mathcal{U}\cup\mathcal{S},\mathcal{E}\right)$, an augmenting path is an alternating path that ends in an unsaturated vertex of $\mathcal{S}$. Here, an alternating path with respect to $\mathcal{M}$ is defined as a path in $\mathcal{G}$ which starts in $\mathcal{U}$ at an unsaturated vertex and then contains, alternately, edges from $\mathcal{E}\setminus\mathcal{M}$ and from $\mathcal{M}$. Therefore, the basic principle of the Hopcroft-Karp algorithm is to search the shortest augmenting path from all of the unsaturated vertices in $\mathcal{S}$ simultaneously \cite{Hopcroft1973}. If there are augmenting paths, a new matching with more saturated vertices can be obtained by computing the symmetric difference between the previous matching and the augmenting paths. It can be shown that the Hopcroft-Karp algorithm is capable of finding a maximum matching for any bipartite graph with the time complexity of $\mathcal{O}\left(N^{2.5}\right)$, which is the fastest deterministic sequential matching algorithm ever known.

The Hopcroft-Karp algorithm, however, can only compute the maximum matching. It is still needed to be generalized for generating the maximum $f$-matching. For a sample of the RBG formulation of the considered multi-carrier multi-access channel, say $\mathcal{G}\left(\mathcal{U}\cup\mathcal{S},\mathcal{E}\right)$, a new bipartite graph $\mathcal{G}(\mathcal{\widetilde{U}}\cup\mathcal{S},\widetilde{\mathcal{E}})$ will then be constructed as follows: For the user $u_{m}\in\mathcal{U}$, let $\mathcal{X}_{u_{m}}$ be a set of $K_{m}$ elements such that $\mathcal{X}_{u_{m}}$ and $\mathcal{X}_{u_{m'}}$ are disjointed if $m\neq m'$, i.e., expand vertex $u_{m}$ to a set of $K_{m}$ vertices. Let
\begin{equation}\label{eq:vertex_extension}
	\widetilde{\mathcal{U}}=\bigcup_{u_{m}\in\mathcal{U}}\mathcal{X}_{u_{m}},
\end{equation}
then connect each element in $\mathcal{X}_{u_{m}}$ to each element of $\mathcal{S}$, whenever $u_{m}$ and $s_{n}$ are adjacent in $\mathcal{G}\left(\mathcal{U}\cup\mathcal{S},\mathcal{E}\right)$. The induced new edge set is denoted by $\widetilde{\mathcal{E}}$

\medskip
\begin{lem}\label{lem:perfect_f_matching}
The bipartite graph $\mathcal{G}(\mathcal{\widetilde{U}}\cup\mathcal{S},\widetilde{\mathcal{E}})$ has a perfect matching if and only if $\mathcal{G}\left(\mathcal{U}\cup\mathcal{S},\mathcal{E}\right)$ has a perfect $f$-matching.
\end{lem}
\begin{IEEEproof}
See Appendix \ref{app:perfect_f_matching}.
\end{IEEEproof}
\medskip

From Lemma \ref{lem:perfect_f_matching}, we only need to extend the user vertices as Eq. \eqref{eq:vertex_extension}. The maximum $f$-matching can then be generated by using Hopcroft-Karp algorithm. However, the maximum $f$-matching method also requires Eq. \eqref{eq:opt_cond_fmatching} to be hold for achieving the optimality of the framework. Moreover, Hopcroft-Karp algorithm will always saturate the user vertices in sequence, which is not fair for the users in subchannel allocation. Therefore, the vertices in the extended set $\widetilde{\mathcal{U}}$ should be grouped and random rotated with the following rules: The vertex set $\widetilde{\mathcal{U}}$ will be divided into $K_{(1)}$ subsets, where the $i$th subset is denoted by $\widetilde{\mathcal{U}}_{i},\,i=1,\ldots,K_{(1)}$. All of the vertices in the set $\mathcal{X}_{u_{m}}$ will be distributed uniformly into these subsets. Thus, the subset $\widetilde{\mathcal{U}}_{i}$ has $\left\lceil\frac{K_{m}}{K_{(1)}}\right\rceil$ vertices from the set $\mathcal{X}_{u_{m}}$ for any $u_{m}\in\mathcal{U}$. Clearly, $\widetilde{\mathcal{U}}_{i}\cap\widetilde{\mathcal{U}}_{i'}=\emptyset,\,\forall i'\neq i$, and $\bigcup_{i=1}^{K_{(1)}}\widetilde{\mathcal{U}}_{i}=\widetilde{\mathcal{U}}$. In the ideal case, all of the subsets $\widetilde{\mathcal{U}}_{i}$ have the identical number of elements from the same set $\mathcal{X}_{u_{m}}$. To guarantee the fairness among all of the users, the family of the subsets $\widetilde{\mathcal{U}}_{i},\,i=1,\ldots,K_{(1)}$ will be randomly rotated, and all of the vertices in each subset $\widetilde{\mathcal{U}}_{i}$ will also be randomly rotated.

With the development of multi-core processors, the parallel algorithms, which can execute on multiple processors simultaneously, attract much more attentions from both industry and academia. In order to fulfill the stringent computation time requirement of the next generation communication systems, the parallel implementation of Hopcroft-Karp algorithm will also be applied in the considered multi-carrier multi-access. Hence, the proposed algorithm will be referred to as the \emph{parallel vertices extension \& random rotation based Hopcroft-Karp} (PVER\textsuperscript{2}HK) algorithm.

\begin{algorithm}[t]
\begin{algorithmic}[1]
	\STATE Input $\mathcal{G}\left(\mathcal{U}\cup\mathcal{S},\mathcal{E}\right)$ and let $l^{*}\leftarrow2\log^{\eta}N+1$.
	\STATE Expand the bipartite graph $\mathcal{G}\left(\mathcal{U}\cup\mathcal{S},\mathcal{E}\right)$ to $\widetilde{\mathcal{G}}(\widetilde{\mathcal{U}}\cup\mathcal{S},\widetilde{\mathcal{E}})$.
	\STATE Generated $(\widetilde{\mathcal{U}}_{i})_{i=1}^{K_{(1)}}$ from $\widetilde{\mathcal{U}}$.
	\STATE Randomly rotate the vertex sets $(\widetilde{\mathcal{U}}_{i})_{i=1}^{K_{(1)}}$ and the elements in them.
	\STATE Let $\widetilde{\mathcal{M}}\leftarrow\emptyset$ and $l\leftarrow0$.
	\REPEAT
		\STATE $\mathcal{G}_{\mathrm{d}}\leftarrow\mathrm{D}(\widetilde{\mathcal{G}},\widetilde{\mathcal{M}})$: Construct a digraph $\mathcal{G}_{\mathrm{d}}$ form $\widetilde{\mathcal{G}}$ with $\widetilde{\mathcal{M}}$, and define $\widetilde{\mathcal{U}}'$ and $\mathcal{S}'$ as the sets of free vertices in $\widetilde{\mathcal{U}}$ and $\mathcal{S}$, respectively.
		\STATE $\left(\mathcal{L},l\right)\leftarrow\mathrm{PBFS}(\widetilde{\mathcal{U}}',\mathcal{S}',\mathcal{G}_{\mathrm{d}},l^{*})$: Execute a parallel breadth-first searching to find the $l$-layer graph $\mathcal{L}$ with $l<l^{*}$. If $\mathcal{L}$ has more layers than $l^{*}$, we let $l=\infty$.
		\STATE $\mathcal{P}\leftarrow\mathrm{PVDP}\left(\mathcal{L},l\right)$: Search in parallel a maximal set of vertex-disjoint paths $\mathcal{P}$ from $\widetilde{\mathcal{U}}'$ to $\mathcal{S}'$ of length $l$ in the layer graph $\mathcal{L}$.
		\STATE $\mathcal{M}\leftarrow\mathrm{PA}(\mathcal{P},\widetilde{\mathcal{M}})$: Execute the parallel augmenting operation to get the next matching.
	\UNTIL{$\mathcal{P}=\emptyset$ or $l>\eta$.}
	\STATE Map $\widetilde{\mathcal{M}}$ to $\mathcal{M}_{f}^{\mathrm{m}}$ by applying the reversion of the vertex expansion and random rotation process.
	\STATE Construct $\left(\mathcal{S}_{m}\right)_{m=1}^{M}$ from $\mathcal{M}_{f}^{\mathrm{m}}$, and compute $\mathcal{S}_{\mathrm{alloc}}^{\mathrm{c}}=\mathcal{S}\setminus\bigcup_{u_{m}\in\mathcal{U}}\mathcal{S}_{m}$.
	\STATE $\left(\mathcal{S}_{m}\right)_{m=1}^{M}\leftarrow\mathrm{RA}\left(\mathcal{S}_{\mathrm{alloc}}^{\mathrm{c}}\right)$: Randomly allocate $\widetilde{K}_{m}-k_{m}$ outage subchannels in $\mathcal{S}_{\mathrm{alloc}}^{\mathrm{c}}$ to the user $u_{m}$.
	\STATE Output subchannel allocation results $\left(\mathcal{S}_{m}\right)_{m=1}^{M}$, then stop.
\end{algorithmic}
\caption{PVER\textsuperscript{2}HK}\label{alg:PVER2HK}
\end{algorithm}

In \cite{Karpinski1998}, a parallel implementation of Hopcroft-Karp algorithm is studied, which can find the maximum matching with a poly-logarithmic complexity within a given approximation error. A matching $\mathcal{M}$ is said to approximate a maximum matching $\mathcal{M}^{\mathrm{m}}$ with an approximation factor $\epsilon$ if and only if
\begin{equation}
	\left|\mathcal{M}\right|\geq\left(1-\epsilon\right)\left|\mathcal{M}^{\mathrm{m}}\right|.
\end{equation}
The detailed description of PVER\textsuperscript{2}HK is listed in Algorithm \ref{alg:PVER2HK}. For the operation $\mathrm{D}(\widetilde{\mathcal{G}},\mathcal{M})$, we orient all the matching edges in $\mathcal{M}$ from their ends in $\widetilde{\mathcal{U}}$ to the ends in $\mathcal{S}$, while the other edges have an opposite direction, i.e., from $\mathcal{S}$ to $\widetilde{\mathcal{U}}$. The operation $\mathrm{PBFS}(\widetilde{\mathcal{U}},\mathcal{S},\mathcal{G}_{\mathrm{d}},l^{*})$ will start at all the vertices in $\widetilde{\mathcal{U}}$ and search along the directed edges in $\mathcal{G}_{\mathrm{d}}$, and end when some of them first hit on the vertices in $\mathcal{S}$. Then, the layer graph $\mathcal{L}$ can be generated with $l<l^{*}$. The $\mathrm{PVDP}\left(\mathcal{L},l\right)$ operation executes the parallel maximal matching algorithm between two vertex layers in order to search all of the vertex-disjoint paths, which are just all the augmenting paths with length $l$. The operation $\mathrm{PA}\left(\mathcal{P},\mathcal{M}\right)$ will execute the augmenting operation on $\mathcal{M}$, which yields a larger maximum matching. All of the parallel operations will work on multiple processors simultaneously. The performance of this parallel implementation is summarized in the following theorem.

\medskip
\begin{thm}
Algorithm \ref{alg:PVER2HK} generates a matching in bipartite graphs with an approximation factor $\log^{-\eta}N$ for some constant $\eta$. The time complexity is $\mathcal{O}\left(\log^{2\eta}N\right)$.
\end{thm}
\medskip

This assertion is the direct consequence of Lemma 12.1.1 and Theorem 12.3.2 in \cite{Karpinski1998}. It can be seen that the proposed algorithm is even faster than FFT, $\mathcal{O}\left(N\log N\right)$, which is essentially a parallel implementation of discrete Fourier transform (DFT).

\subsection{Asymptotic Optimal Coding Schemes}\label{subsec:coding_schemes}

As stated before, the optimal outage performance can only be achieved by using the maximum $f$-matching method with the help of coding schemes. To the best of our knowledge, however, the optimal coding scheme discussed in Section \ref{sec:coding} is a NP-Hard problem and still unknown by present. However, there are two main practical approaches to maximize the minimum product distance. The induced coding schemes are capable of achieving the optimal DMT in parallel fading channels, i.e., the asymptotic optimal coding schemes.

The first class is the rotated $\mathbb{Z}^{\widetilde{K}_{m}}$-lattices code for the user $u_{m}$, which is based on the algebraic number theory and lattices theory. The basic idea is to construct the $\widetilde{K}_{m}$-dimension constellation with the size $2^{R_{m}}$ based on $\mathbb{Z}^{\widetilde{K}_{m}}$-lattices. Then, the constellation will be rotated to an appropriate angle such that $D_{p}^{\min}$ is maximized. According to the lattices theory, $D_{p}^{\min}$ can be calculated in theory. One can refer to \cite{Wang2003,Oggier2004} and its references for detailed discussions on this coding scheme. Another advantage of the rotated $\mathbb{Z}^{\widetilde{K}_{m}}$-lattices code lies in the fact that it can reduce the peak-to-average power ratio (PAPR) \cite{Henkel2000}. As a matter of fact, SC-FDMA can be seen as a special case of rotated $\mathbb{Z}^{\widetilde{K}_{m}}$-lattices codes, where the rotated angle is determined by the DFT matrix. This rotation increases $D_{p}^{\mathrm{min}}$ or the distance between two points with zero product distance. Therefore, compared to uncoded OFDMA systems, SC-FDMA has a better performance in form of decoding error probability and PAPR.

Another approach is the permutation code which was proposed to achieve the optimal DMT \cite{Tse2005,Tavildar2006}. The basic idea is to use the constellation of size $2^{R_{m}}$ on a complex plane for each subchannel. The points in the constellation for each subchannel are permuted, so that the product distance is maximized. The permutation operation on each subchannel can be seen as that of the original information times a specific matrix, which is referred to as the universal decodable matrix (UDM). Based on the Pascal triangle, the UDMs can be constructed directly for every subchannel \cite{Ganesan2007}. From the perspective of rotated $\mathbb{Z}^{\widetilde{K}_{m}}$-lattices code, the permutation code can be seen as a way to construct a $\widetilde{K}_{m}$-dimension constellation with size $2^{\widetilde{K}_{m}R_{m}}$. The permutation rules imply that we should choose $2^{R_{m}}$ points from $2^{\widetilde{K}_{m}R_{m}}$, so that the product distance is maximized. It is not trivial to evaluate the decoding error probability of the permutation code, because the analytic formula for $D_{p}^{\min}$ has not been obtained. According to the construction process of the permutation code, the PAPR performance could be worse than the rotated $\mathbb{Z}^{\widetilde{K}_{m}}$-lattices code for the same average power. Another important difference is that the constellation size of each subchannel is $2^{R_{\mathrm{s}}}=2^{R_{m}/\widetilde{K}_{m}}$ for the rotated $\mathbb{Z}^{\widetilde{K}_{m}}$-lattices code, while it is $2^{R_{m}}$ for the permutation code.

\section{Simulation Results}\label{sec:simulation_results}

In this section, some simulation examples are presented to verify the theoretical derivations and the proposed algorithm. In the first group of simulation results, the tight upper bound of the conditional outage probability is verified, which is the key fundamental of the proposed code $f$-matching framework. The second group of simulation results verify the outage performance of the proposed framework.

For the first group of simulation results, a large number of samples are generated at each SNR value according to the channel model in Section \ref{subsec:channel_model}. The number of outage events can be obtained under the condition that the transmitter knows $\unit[1]{bit}$ CSI by computing the instantaneous channel capacity. The conditional outage probability $p^{\mathrm{cop}}_{m}\left(R_{m}\left|\mathcal{D}_{K_{m}k_{m}}\right.\right)$ can then be calculated accordingly. The theoretical approximations are calculated by applying the results in Theorem \ref{thm:con_outage_exponent}.

\begin{figure}[t]
	\centering
	\includegraphics[width=3.6in]{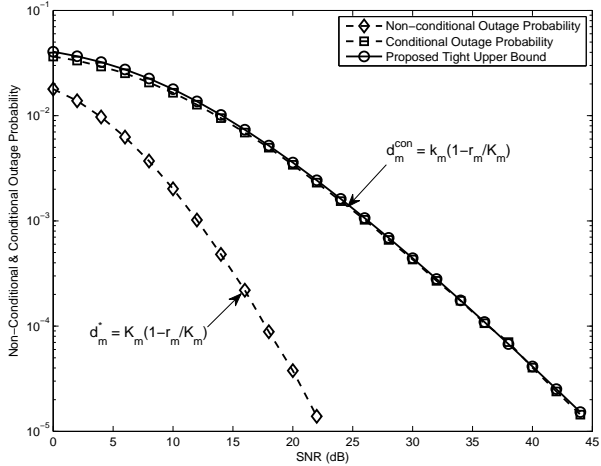}
	\caption{The tight upper bound of the conditional outage probability for the user $u_{m}$ with $K_{m}=4$, $k_{m}=2$, and $r_{m}=1.2$.}\label{fig:cond_outage_exponent}
\end{figure}

\begin{figure}[t]
	\centering
	\includegraphics[width=3.6in]{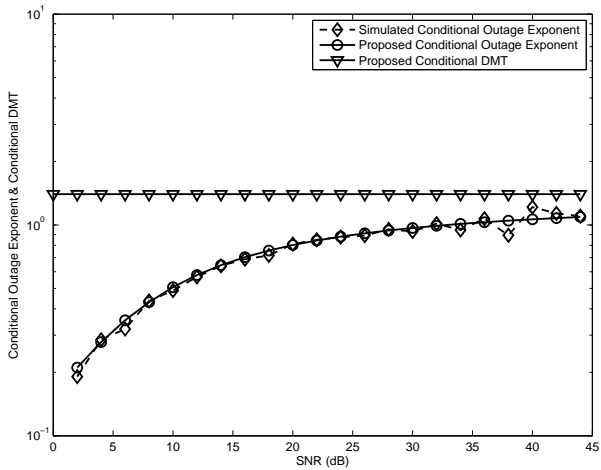}
	\caption{The conditional outage exponent and conditional DMT for the user $u_{m}$ with $K_{m}=4$, $k_{m}=2$, and $r_{m}=1.2$.}\label{fig:diversity_gain}
\end{figure}

Fig. \ref{fig:cond_outage_exponent} compares the simulation results and the theoretical curves of the conditional outage probability. The user $u_{m}$ has $K_{m}=4$ coherence bandwidths with $k_{m}=2$ non-outage ones, and the multiplexing gain $r$ is set to $1.2$. It can be seen that the proposed tight upper bound is nearly identical with the simulation results in the realistic SNR range. For comparison, the curve of non-conditional outage probability is also plotted. The outage performance with $\unit[1]{bit}$ CSI is worse than the corresponding one without CSI in this situation. In the high SNR regime, the slope of the conditional outage probability is determined by Eq. \eqref{eq:dmt}, while the slope of the outage probability is determined by $K_{m}\left(1-\frac{r_{m}}{K_{m}}\right)$. Fig. \ref{fig:diversity_gain} presents the curves of conditional outage exponent and conditional DMT at a given multiplexing gain. Clearly, the conditional outage exponent is an increasing function of SNR for a fixed multiplexing gain. The proposed theoretical curve is nearly the same as the simulation one. In Fig. \ref{fig:diversity_gain}, the conditional DMT is plotted as a constant which is much larger than the outage exponent at realistic SNRs. However, the conditional outage exponent is approaching the conditional DMT as SNR tends to infinity. Therefore, the proposed conditional outage exponent can be used to estimate the decreasing slope of conditional outage probabilities for the considered multi-carrier multi-access with $\unit[1]{bit}$ CSI.

For the second group of simulation results, a large number of samples are generated according to the RBG formulation $\mathscr{G}\left\{\mathcal{K}_{MN},\mathsf{P}\right\}$ of the considered multi-carrier multi-access in Section \ref{subsec:RBG_Formulation}. The subchannels are then allocated by the proposed PVER\textsuperscript{2}HK algorithm to each user for RB and chunk based coded $f$-matching allocation schemes, respectively. The simulation values of the outage probability can then be obtained by counting the number of outage events of each user in the total number of simulations. In proposed allocation schemes, the first order approximation of the outage probabilities are calculated by the formulas in Theorem \ref{thm:outage_RB} and Theorem \ref{thm:outage_chunk}, respectively. Besides, the interleaved allocation scheme, TDMA based scheme, and the localized allocation scheme have been simulated so as to illustrate the performance gain of the coded $f$-matching approach. The asymptotic line $\mathsf{SNR}^{d_{m}^{*}\left(r_{m}\right)}$ is also plotted to compare the slope of these curves in high SNRs.

\begin{figure}[t]
	\centering
	\includegraphics[width=3.6in]{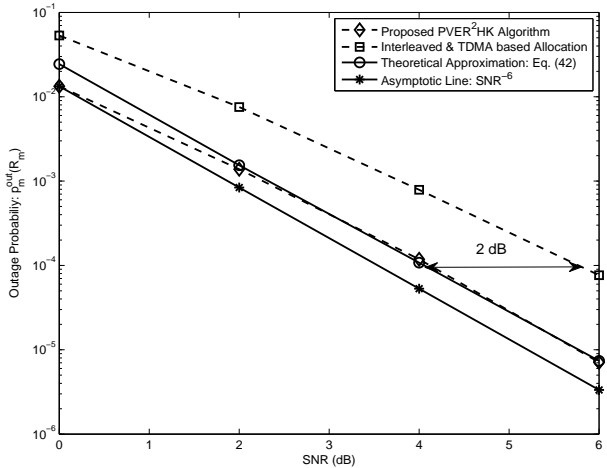}
	\caption{The outage probability of the RB based coded $f$-matching allocation scheme for the user $u_{m}$ with $K_{m}=3$ and $R_{m}=1$ ($r_{m}=0$).}\label{fig:RB_1}
\end{figure}

\begin{figure}[t]
	\centering
	\includegraphics[width=3.6in]{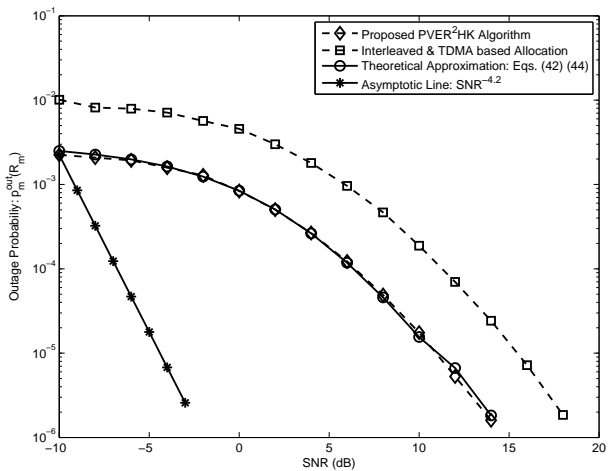}
	\caption{The outage probability of the RB based coded $f$-matching allocation scheme for the user $u_{m}$ with $K_{m}=3$ and $r_{m}=0.9$.}\label{fig:RB_2}
\end{figure}

Figs. \ref{fig:RB_1} and \ref{fig:RB_2} compare the simulation results and the theoretical curves of the RB based coded $f$-matching allocation scheme. These simulation examples consider the multi-carrier multi-access channel with $M=2$ users and $L=6$ coherence bandwidths, where each coherence bandwidth contains $N_{\mathrm{c}}=M$ RBs. The fixed target rate schemes, i.e., $R=1$, are plotted in Fig. \ref{fig:RB_1}. It can be seen that both PVER\textsuperscript{2}HK algorithm and the interleaved \& TDMA based allocation scheme have the same slope in the high SNR regime, which means that they have the same diversity order. The proposed algorithm, however, has $\unit[2]{dB}$ SNR gain. The theoretical approximation in Eq. \eqref{eq:outage_RB_high} approaches the simulation results as SNR increases. In Fig. \ref{fig:RB_2}, the dynamic rate scenario is considered, where the multiplexing gain $r$ is set to $0.9$. The theoretical approximation is computed by Eq. \eqref{eq:outage_RB_high} and Eq. \eqref{eq:outage_RB_LowSNR} in high and low SNR regimes, respectively. It can be seen that the theoretical approximation is nearly identical with the simulation results of the proposed PVER\textsuperscript{2}HK algorithm. Compared with the interleaved \& TDMA based allocation schemes, the proposed algorithm has $\unit[4]{dB}$ performance gain. Therefore, the performance gain of the proposed algorithm increases as the multiplexing gain increases.

\begin{figure}[t]
	\centering
	\includegraphics[width=3.6in]{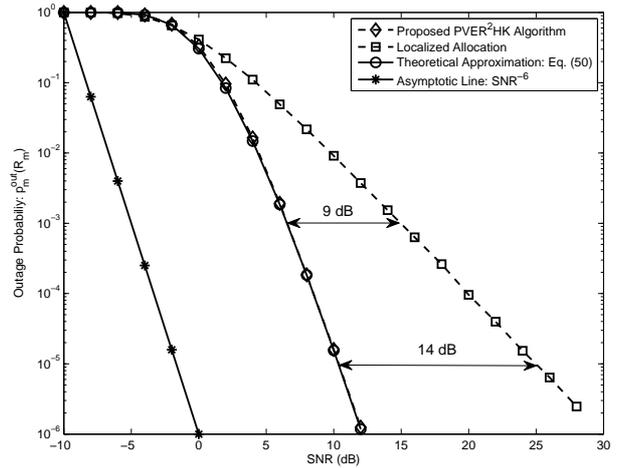}
	\caption{The outage probability of the chunk based coded $f$-matching allocation scheme for the user $u_{m}$ with $K_{m}=1$ and $R_{m}=1$ ($r_{m}=0$).}\label{fig:chunk_1}
\end{figure}

\begin{figure}[t]
	\centering
	\includegraphics[width=3.6in]{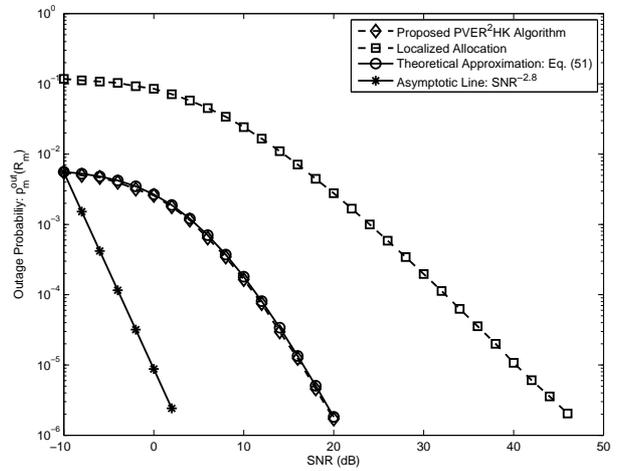}
	\caption{The outage probability of the chunk based coded $f$-matching allocation scheme for the user $u_{m}$ with $K_{m}=2$ and $r_{m}=0.6$.}\label{fig:chunk_2}
\end{figure}

Figs. \ref{fig:chunk_1} and \ref{fig:chunk_2} compare the simulation results and the theoretical curves of chunk based coded $f$-matching allocation schemes. These simulation examples consider the multi-carrier multi-access channel with $M=3$ users and $L=6$ coherence bandwidths, where $N_{\mathrm{c}}$ RBs in one coherence bandwidth is bundled as one chunk. The fixed target rate schemes, i.e., $R=1$, are plotted in Fig. \ref{fig:chunk_1}. The PVER\textsuperscript{2}HK algorithm has the full diversity order, whereas the diversity gain of the localized allocation scheme is $2$. Therefore, the SNR gain of the proposed algorithm increases as SNR increases. The theoretical approximation in Eq. \eqref{eq:outage_chunk_1} is nearly identical with the simulation results of the proposed algorithm. The dynamic rate scenario is presented in Fig. \ref{fig:chunk_2}, where the multiplexing gain $r$ is set to $0.6$. The theoretical approximation is computed by Eq. \eqref{eq:outage_chunk_2}. Once again, the theoretical approximation is nearly identical with the simulation result of the proposed PVER\textsuperscript{2}HK algorithm. Compared with the fixed rate scenario, the proposed algorithm has greater SNR gains, i.e., the performance gain of the proposed algorithm increases as the multiplexing gain increases.

\section{Conclusions}\label{sec:conclusions}
This paper proposed an analytic coded $f$-matching framework for channel resources allocation in multi-carrier multi-access channels, where the OER and DMR were defined as the comprehensive performance metrics. Compared to conventional approaches, the proposed framework sufficiently combined the advantages of optimal frequency-domain coding scheme and RBG based maximum $f$-matching approach, which can fully explore the combinatorial structure of the channel allocation problem. It was shown in theory that the outage probability, OER and DMR are optimal such that all of the users share the total multiplexing gain while achieving the full frequency diversity. The low complexity optimal allocation algorithm is a natural result of the proposed framework, which only depends on $\unit[1]{bit}$ CSI feedback. The simulation results did not only verify the theoretical derivations, but also showed the significant performance gains in both diversity and SNR. Therefore, the proposed framework provides powerful tools for designing future multi-carrier multi-access systems, and has the potential to generalized to other wireless communication systems.

% or


\appendices

\section{Proof of Theorem \ref{thm:con_outage_exponent}}\label{app:con_outage_exponent}

Define a sequence of random variables $\left\{Z_{mn}\right\}_{n=1}^{K_{m}}$ with the following distribution
\begin{equation}
\begin{aligned}
	& F_{Z_{mn}}\left(z\right)=\Pr\left\{\left.I_{mn}<z\right|\mathcal{D}_{K_{m}k_{m}}\right\}\\
 	& =\left\{
 	\begin{aligned}
		& 1-\frac{1}{q_{\mathrm{s}}}\exp\left(-\frac{e^{N_{\mathrm{c}}z}-1}{\gamma}\right), & n=1,\ldots,k_{m};\\
		& \frac{1}{p_{\mathrm{s}}}\left[1-\exp\left(-\frac{e^{N_{\mathrm{c}}z}-1}{\gamma}\right)\right], & n=k_{m}+1,\ldots,K_{m}.
	\end{aligned}\right.
\end{aligned}
\end{equation}
Let $X_{n}=R_{\mathrm{s}}-Z_{mn}$ and $Y_{K_{m}}=\sum_{n=1}^{K_{m}}X_{n}$, then the conditional outage probability defined in Eq. \eqref{eq:con_outage_probability} is given by
\begin{equation}
\begin{aligned}
	p_{m}^{\mathrm{cop}}\left(R_{m}\left|\mathcal{D}_{K_{m}k_{m}}\right.\right) & =\Pr\left\{\left.\sum_{n=1}^{K_{m}}I_{mn}<R_{m}\right|\mathcal{D}_{K_{m}k_{m}}\right\}\\
	& =\Pr\left\{Y_{K_{m}}>0\right\}.
\end{aligned}
\end{equation}

According to the considered condition, the elements of $\left\{h_{mn}\right\}_{n=1}^{K_{m}}$ are independent with the identical distribution of $\mathcal{CN}\left(0,1\right)$. Let $1\leq n_{1}\leq k_{m}$, $k_{m}+1\leq n_{2}\leq K_{m}$, the cumulant-generating function of $Y_{K_{m}}$ is then given by
\begin{equation}
\begin{aligned}
	& \Phi_{K_{m}}\left(\lambda\right)=\ln\mathsf{E}\left\{e^{\lambda}Y_{K_{m}}\right\}=K_{m}R_{\mathrm{s}}\lambda\\
	& +\ln\left(\mathsf{E}\left\{e^{-\lambda Z_{mn_{2}}}\right\}\right)^{\beta_{m}K_{m}}+\ln\left(\mathsf{E}\left\{e^{-\lambda Z_{mn_{1}}}\right\}\right)^{\left(1-\beta_{m}\right)K_{m}}\\
	& =K_{m}\left[R_{\mathrm{s}}\lambda+\beta_{m}\ln\int_0^{R_{\mathrm{s}}}e^{-\lambda z}dF_{Z_{mn_{2}}}\left(z\right)\right.\\
	& \left.+\left(1-\beta_{m}\right)\ln\int_{R_{\mathrm{s}}}^{\infty}e^{-\lambda z}dF_{Z_{mn_{1}}}\left(z\right)\right]\\
	& =K_{m}\left[\beta_{m}\ln\int_0^{R_{\mathrm{s}}}\frac{N_{\mathrm{c}}e^{-\lambda z}}{p_\mathrm{s}\gamma}\exp\left(-\frac{e^{N_{\mathrm{c}}z}-1}{\gamma}+N_{\mathrm{c}}z\right)dz\right.\\
	& \left.+\left(1-\beta_{m}\right)\ln\int_{R_{\mathrm{s}}}^{\infty}\frac{N_{\mathrm{c}}e^{-\lambda z}}{q_{\mathrm{s}}\gamma}\exp\left(-\frac{e^{N_{\mathrm{c}}z}-1}{\gamma}+N_{\mathrm{c}}z\right)dz\right]\\
    &+K_{m}R_{\mathrm{s}}\lambda\\
    & =K_{m}\left[R_{\mathrm{s}}\lambda+\beta_{m}\ln\frac{e^{\frac{1}{\gamma}}}{p_\mathrm{s}\gamma^{\frac{\lambda}{N_{\mathrm{c}}}}}+\beta_{m}\ln\int_{\frac{1}{\gamma}}^{\frac{e^{R_\mathrm{c}}}{\gamma}}t^{-\frac{\lambda}{N_{\mathrm{c}}}}e^{-t}dt\right.\\
    &\left.+\left(1-\beta_{m}\right)\ln\frac{e^{\frac{1}{\gamma}}}{q_\mathrm{s}\gamma^{\frac{\lambda}{N_{\mathrm{c}}}}}+\left(1-\beta_{m}\right)\ln\int_{\frac{e^{R_{\mathrm{c}}}}{\gamma }}^{\infty }t^{-\frac{\lambda}{N_{\mathrm{c}}}}e^{-t}dt\right]
\end{aligned}
\end{equation}
\begin{equation*}
\begin{aligned}
	& =K_{m}\left[\left(R_{\mathrm{c}}-\ln\gamma\right)\frac{\lambda}{N_{\mathrm{c}}}+\frac{1}{\gamma}-\ln p_{\mathrm{s}}^{\beta_{m}}q_{\mathrm{s}}^{1-\beta_{m}}\right.\\
	& +\beta_{m}\ln\left(\Gamma\left(1-\frac{\lambda}{N_{\mathrm{c}}},\frac{1}{\gamma}\right)-\Gamma\left(1-\frac{\lambda}{N_{\mathrm{c}}},\frac{e^{R_{\mathrm{c}}}}{\gamma}\right)\right)\\
	& \left.+\left(1-\beta_{m}\right)\ln\Gamma\left(1-\frac{\lambda}{N_{\mathrm{c}}},\frac{e^{R_{\mathrm{c}}}}{\gamma}\right)\right],
\end{aligned}
\end{equation*}
where $\beta_{m}=1-\frac{k_{m}}{K_{m}}$. Therefore, we have
\begin{equation}
\begin{aligned}
	\Lambda\left(\lambda\right)=&\lim_{K_{m}\rightarrow\infty}\frac{1}{K_{m}}\Phi_{K_{m}}\left(\lambda\right)\\
	= & \left(R_{\mathrm{c}}-\ln\gamma\right)\lambda+\frac{1}{\gamma}-\ln p_{\mathrm{s}}^{\beta_{m}}q_{\mathrm{s}}^{1-\beta_{m}}\\
	& +\beta_{m}\ln\left(\Gamma\left(1-\lambda,\frac{1}{\gamma}\right)-\Gamma\left(1-\lambda,\frac{e^{R_{\mathrm{c}}}}{\gamma}\right)\right)\\
	& +\left(1-\beta_{m}\right)\ln\Gamma\left(1-\lambda,\frac{e^{R_{\mathrm{c}}}}{\gamma}\right).
\end{aligned}
\end{equation}
Considering the relationship between the cumulant-generating function and the characteristic function, the characteristic function of $Y_{K_{m}}$ can then by given by
\begin{equation}
	\Phi_{K_{m}}\left(i\lambda\right)=K_{m}\Lambda\left(i\lambda\right).
\end{equation}

Define $G\left(y\right)=\Pr\left\{Y_{K_{m}}>y\right\}$, then according to L\'evy's theorem \cite{Shiryaev1996}, we have
\begin{equation}
	\begin{aligned}
		G\left(y\right) & =\frac{1}{2\pi i}\int_{-\infty}^{+\infty}\frac{1}{\xi}\exp\left(K_{m}\Lambda\left(i\xi\right)-i\xi y\right)d\xi\\
		& =\frac{1}{2\pi i}\int_{\lambda-i\infty}^{\lambda+i\infty}\frac{1}{z}\exp\left(K_{m}\Lambda\left(z\right)-zy\right)dz,
	\end{aligned}
\end{equation}
where $z=\lambda+i\xi$, and $\lambda$ is chosen from the convergence region of this integral. According to the saddle-point approximation method \cite{Butler2007}, if we let $z^{*}$ with $\Re\left(z^{*}\right)=\lambda^{*}$ be a solution of the saddle-point equation
\begin{equation}
	\Lambda'\left(z\right)=\frac{y}{K_{m}},
\end{equation}
then
\begin{equation}
	\begin{aligned}
		& \frac{1}{2\pi i}\int_{\lambda^{*}-i\infty}^{\lambda^{*}+i\infty}\frac{1}{z}\exp\left(K_{m}\Lambda\left(z\right)-zy\right)dz\lesssim\\
		& \frac{1}{\sqrt{2\pi K_{m}\Lambda''\left(\lambda^{*}\right)}\lambda^{*}}\exp\left(K_{m}\Lambda\left(z^{*}\right)-z^{*}y\right).
	\end{aligned}
\end{equation}
Since $G\left(y\right)$ is a real function, we can always choose a real $z^{*}$ only if the cumulant-generating function exists. Moreover, $\frac{\lambda y}{K_{m}}-\Lambda\left(\lambda\right)$ is a conjugate function when $\lambda\in\mathbb{R}$, then $\lambda^{*}$ must be unique. Therefore, the exponentially tight upper bound of the conditional outage probability $p_{m}^{\mathrm{cop}}\left(R_{m}\left|\mathcal{D}_{K_{m}k_{m}}\right.\right)$ is the tail distribution of $G\left(y\right)$ with $y>\mathsf{E}\left\{Y_{K_{m}}\right\}$, which is given by
\begin{equation}\label{eq:saddle-point}
\begin{aligned}
	& p_{m}^{\mathrm{cop}}\left(R_{m}\left|\mathcal{D}_{K_{m}k_{m}}\right.\right)=G\left(0\right)\\
	\lesssim & \frac{1}{\sqrt{2\pi K_{m}\sigma^{2}}\lambda^{*}}\exp\left(K_{m}\Lambda\left(\lambda^{*}\right)\right)=p_{m}^{\mathrm{upper}}\left(R_{m}\left|\mathcal{D}_{K_{m}k_{m}}\right.\right),
\end{aligned}
\end{equation}
where $\lambda^{*}$ is the solution of $\Lambda'\left(\lambda\right)=0$, and $\sigma^{2}=\Lambda''\left(\lambda^{*}\right)$. Clearly, we have
\begin{equation}\label{eq:CGF_derivative}
\begin{aligned}
	\Lambda'\left(\lambda\right)= & \left(R_{\mathrm{c}}-\ln\gamma\right)+\left(1-\beta_{m}\right)\frac{\Gamma'\left(1-\lambda,\frac{e^{R_{\mathrm{c}}}}{\gamma}\right)}{\Gamma\left(1-\lambda,\frac{e^{R_{\mathrm{c}}}}{\gamma}\right)}\\
	& +\beta_{m}\frac{\Gamma'\left(1-\lambda,\frac{1}{\gamma}\right)-\Gamma'\left(1-\lambda,\frac{e^{R_{\mathrm{c}}}}{\gamma}\right)}{\Gamma\left(1-\lambda,\frac{1}{\gamma}\right)-\Gamma\left(1-\lambda,\frac{e^{R_{\mathrm{c}}}}{\gamma}\right)}=0.
\end{aligned}
\end{equation}
According to the properties of Gamma function and Meijer's $G$-function, we have
\begin{equation}\label{eq:gamma_derivative}
\begin{aligned}
	\Gamma'\left(1-\lambda,z\right)= & -\Gamma\left(1-\lambda,z\right)\ln z\\
	& -G_{2,3}^{3,0}\left(z\left|
	\begin{array}{ccc}
		1, & 1, & -\\
		0, & 0, & 1-\lambda
	\end{array}\right.
	\right).
\end{aligned}
\end{equation}
Therefore, Eq. \eqref{eq:exponent_equation} can be obtained by plugging Eq. \eqref{eq:gamma_derivative} into Eq. \eqref{eq:CGF_derivative}. For $\sigma^{2}$, we have
\begin{equation}
\begin{aligned}\label{eq:CGF_parameter}
	& \sigma^{2}=\Lambda''\left(\lambda^{*}\right)\\
	=&\left(1-\beta_{m}\right)\left[\frac{\Gamma''\left(1-\lambda^{*},\frac{e^{R_{\mathrm{c}}}}{\gamma}\right)}{\Gamma\left(1-\lambda^{*},\frac{e^{R_{\mathrm{c}}}}{\gamma}\right)}-\left(\frac{\Gamma'\left(1-\lambda^{*},\frac{e^{R_{\mathrm{c}}}}{\gamma}\right)}{\Gamma\left(1-\lambda^{*},\frac{e^{R_{\mathrm{c}}}}{\gamma}\right)}\right)^{2}\right]\\
	& +\beta_{m}\frac{\Gamma''\left(1-\lambda^{*},\frac{1}{\gamma}\right)-\Gamma''\left(1-\lambda^{*},\frac{e^{R_{\mathrm{c}}}}{\gamma}\right)}{\Gamma\left(1-\lambda^{*},\frac{1}{\gamma}\right)-\Gamma\left(1-\lambda^{*},\frac{e^{R_{\mathrm{c}}}}{\gamma}\right)}\\
	& -\beta_{m}\left(\frac{\Gamma'\left(1-\lambda^{*},\frac{1}{\gamma}\right)-\Gamma'\left(1-\lambda^{*},\frac{e^{R_{\mathrm{c}}}}{\gamma}\right)}{\Gamma\left(1-\lambda^{*},\frac{1}{\gamma}\right)-\Gamma\left(1-\lambda^{*},\frac{e^{R_{\mathrm{c}}}}{\gamma}\right)}\right)^{2}.
\end{aligned}
\end{equation}
According to the properties of Meijer's $G$-function, we have
\begin{equation}\label{eq:Meijer_derivative}
\begin{aligned}
	& \frac{\partial}{\partial x}G_{2,3}^{3,0}\left(z\left|
	\begin{array}{ccc}
		1, & 1, & - \\
		0, & 0, & 1-\lambda^{*}\\
	\end{array}\right.\right)\\
	= & G_{2,3}^{3,0}\left(z\left|
	\begin{array}{ccc}
		1, & 1, & - \\
		0, & 0, & 1-\lambda^{*}\\
	\end{array}\right.\right)\ln z\\
	&+2G_{3,4}^{4,0}\left(z\left|
	\begin{array}{cccc}
		1, & 1, & 1, & -\\
		0, & 0, & 0, & 1-\lambda^{*}\\
	\end{array}\right.\right)
\end{aligned}
\end{equation}
Therefore, Eq. \eqref{eq:exponent_parameter} can be obtained by plugging Eq. \eqref{eq:Meijer_derivative} into Eq. \eqref{eq:CGF_parameter}.

For the tail distribution with $y<\mathsf{E}\left\{Y_{K_{m}}\right\}$, a similar result can be obtained by applying the same process. Therefore, Theorem \ref{thm:con_outage_exponent} has been established.

\section{Proof of Lemma \ref{lem:f_matching}}\label{app:f_matching}

($\Rightarrow$). Suppose $\mathcal{G}\left(\mathcal{U}\cup\mathcal{S},\mathcal{E}\right)$ has a maximum $f$-matching $\mathcal{M}_{f}^{\mathrm{m}}$, which cannot saturates the vertex $u_{m}$. Then, there must be conflicts between $u_{m}$ and other vertices in $\mathcal{U}$. Therefore, we have a subset $\mathcal{X}\subseteq\mathcal{U}$ with $u_{m}\in\mathcal{X}$ satisfying
\begin{equation}
	\sum_{u_{i}\in\mathcal{X}}f\left(u_{i}\right)=\sum_{u_{i}\in\mathcal{X}}K_{i}>\sum_{s_{n}\in\mathcal{N} \left(\mathcal{X}\right)}f\left(s_{n}\right).
\end{equation}

($\Leftarrow$). Orient all edges of $\mathcal{G}\left(\mathcal{U}\cup\mathcal{S},\mathcal{E}\right)$ from $\mathcal{U}$ to $\mathcal{S}$. Add a source $\theta$ joined all vertices in $\mathcal{U}$ and a sink $\zeta$ to which every vertex in $\mathcal{S}$ is joined. This induced directed graph is denoted by $\mathcal{G}_{\mathrm{d}}$. Assign capacities to all of the edges as follows:
\begin{equation}
	\left\{\begin{aligned}
		& c\left(e\right)=f\left(u_{i}\right), & \textrm{if }e\textrm{ is from }\theta\textrm{ to }u_{i}\in\mathcal{U};\\
		& c\left(e\right)=1, & \textrm{on all the other edges}.
	\end{aligned}\right.
\end{equation}
Choose a subset $\mathcal{X}$ of $\mathcal{U}$ with $u_{m}\in\mathcal{X}$ such that Eq. \eqref{eq:f_matching_cond} holds. Let $\mathcal{C}=\mathcal{X}\cup\left\{\theta\right\}$, which is clearly a cut set separating $\theta$ from $\zeta$. The capacity of $\mathcal{C}$ is then given by
\begin{equation}
	\sum_{u_{i}\in\mathcal{U}-\mathcal{X}}f\left(u_{i}\right)+\sum_{s_{n}\in\mathcal{N}\left(\mathcal{X}\right)}f\left(s_{n}\right)<\sum_{u_{i}\in\mathcal{U}}f\left(u_{i}\right).
\end{equation}
According to max-flow min-cut theorem in Lemma \ref{lem:max_flow_min_cut}, the maximum flow $\mathcal{F}^{*}$ from $\theta$ to $\zeta$ of $\mathcal{G}_{\mathrm{d}}$ must be smaller than $\sum_{u_{i}\in\mathcal{U}}f\left(u_{i}\right)$. Moreover, since $u_{m}\in\mathcal{X}\subseteq\mathcal{C}$, the maximum flow $\mathcal{F}^{*}$ cannot achieve the capacities of all the edges from $\theta$ to vertices in $\mathcal{X}$. Thus, greater flow rate can be allocated to the edge from $\theta$ to $u_{i}$ in $\mathcal{U}$ than the edge from $\theta$ to $u_{m}$ when they are conflicts. Therefore, the generated maximum $f$-matching cannot saturate $u_{m}$.

\section{Proof of Lemma \ref{lem:matching_edges}}\label{app:matching_edges}

Let $\mathcal{G}(\mathcal{U}\cup\mathcal{S},\mathcal{E})$ satisfy the conditions in this lemma, and we assume there is an edge set $\mathcal{E}_{0}$ that contains at least one maximum $f$-matching which cannot saturate $u_{m}$. According to Lemma \ref{lem:f_matching}, there must be a subset $\mathcal{X}\subseteq\mathcal{U}$ with the maximum number of elements, which satisfies
\begin{equation}
   \left|\mathcal{N}\left(\mathcal{X}\right)\right|<K^{\mathrm{sum}}\left(\mathcal{X}\right)=\sum_{u_{i}\in\mathcal{X}}K_{i},
\end{equation}
It can be seen that there are $\left|\mathcal{X}\right|\left(K^{\mathrm{sum}}\left(\mathcal{X}\right)-1\right)$ edges from $\mathcal{X}$ to $\mathcal{S}$ at most.

For any $\mathcal{X}$ with $|\mathcal{X}|=1,\ldots,M$, there are at most
\begin{equation}
	\left|\mathcal{X}\right|\left(K^{\mathrm{sum}}\left(\mathcal{X}\right)-1\right)+\left(M-\left|\mathcal{X}\right|\right)N
\end{equation}
edges. Therefore, $\left|\mathcal{E}\right|$ must be equal or greater than
\begin{equation}\label{eq:edge_number}
\begin{aligned}
	& \max_{|\mathcal{X}|=1,\ldots,M}\left\{\left|\mathcal{X}\right|\left(K^{\mathrm{sum}}\left(\mathcal{X}\right)-1\right)+\left(M-\left|\mathcal{X}\right|\right)N+1\right\}\\
	= & \max_{|\mathcal{X}|=1,\ldots,M}\left\{\left|\mathcal{X}\right|\left(K^{\mathrm{sum}}\left(\mathcal{X}\right)-N-1\right)+MN+1\right\}.
\end{aligned}
\end{equation}
Since $K^{\mathrm{sum}}\left(\mathcal{X}\right)$ is an increasing function of $\left|\mathcal{X}\right|$, the maximum value can only be achieved for $\left|\mathcal{X}\right|=1$ or $\left|\mathcal{X}\right|=M$.

Consider first $\mathcal{X}$ has only one vertex such that $\mathcal{X}=\left\{u_{m}\right\}$. There will be at most $\left(M-1\right)N+K_{m}-1$ edges. Thus, if $\left|\mathcal{E}\right|\geq\left(M-1\right)N+K_{m}$ there must be a maximum $f$-matching which can saturate $u_{m}$ for any $\mathcal{E}$. Because $\left(M-1\right)N+K_{m}>\left(M-1\right)N+K_{i}$ with $K_{\left(i\right)}<K_{\left(m\right)}$, all of the vertices $(u_{\left(i\right)})_{i=1}^{m}$ can be saturated by the maximum $f$-matching for any $\mathcal{E}$. Consider then the case that $\mathcal{X}=\mathcal{U}$. There will be at most $M\left(K^{\mathrm{sum}}-1\right)$ edges. Thus, if $\left|\mathcal{E}\right|\geq M\left(K^{\mathrm{sum}}-1\right)+1$ there must be a maximum $f$-matching which can saturate $u_{m}$ for any $\mathcal{E}$. By solving the inequality that
\begin{equation}
	\left(M-1\right)N+K_{m}\geq M\left(K^{\mathrm{sum}}-1\right)+1,
\end{equation}
we can get
\begin{equation}\label{eq:one}
	K_{m}\leq N+1-\frac{M}{M-1}K_{m}^{\mathrm{sum}}.
\end{equation}
If Eq. \eqref{eq:one} holds, the bipartite graph with $\left|\mathcal{X}\right|=1$ has the maximum number of edges. Otherwise, the bipartite graph with $\left|\mathcal{X}\right|=M$ has the maximum number of edges. Therefore, Lemma \ref{lem:matching_edges} has been established.

\section{Proof of Theorem \ref{thm:outage_RB}}\label{app:outage_RB}
According to the second part of Lemma \ref{lem:matching_edges}, $MN-(M\widetilde{K}^{\mathrm{sum}}-M)=M$ indicates the case that there is one isolated coherence bandwidth in $\mathcal{S}$. Recalling Eq. \eqref{eq:opt_cond_fmatching} in the coded $f$-matching approach, thus any user $u_{m}$ will be allocated $\frac{\widetilde{K}_{m}}{L}$ RBs in each coherence bandwidth. In this context, the outage probability of the user $u_{m}$ has the term
\begin{equation}
\begin{aligned}
	& \frac{1}{M}\binom{L}{L-1}p_{m}^{\mathrm{cop}}\left(R_{m}\left|\mathcal{D}_{L,L-1}\right.\right)p_{\mathrm{s}}^{M}q_{\mathrm{s}}^{M\left(L-1\right)}\\
	= & \frac{Lp_{m}^{\mathrm{cop}}\left(R_{m}\left|\mathcal{D}_{L,L-1}\right.\right)p_{\mathrm{s}}^{M}}{M}+\mathsf{O}\left(p_{\mathrm{s}}^{M+L-1}\right).
\end{aligned}
\end{equation}
Clearly, the lowest order of $p_{\mathrm{s}}$ is given by $M+L-1>L$, since $M\geq2$ in the considered multi-carrier multi-access channel. Therefore, this case is the higher order terms in the formula of the outage probability.

According to the first part of Lemma \ref{lem:matching_edges}, there must be $\kappa$ outage coherence bandwidths for the user $u_{m}$, with $\kappa=L-\left\lceil\frac{\widetilde{K}_{m}}{N_{\mathrm{c}}}\right\rceil+1,\ldots,L$, if this user cannot be saturated by the generated maximum $f$-matching. Recalling the definition of maximum $f$-matching, thus all of the RBs in $L-\kappa$ coherence bandwidths must be allocated to the user $u_{m}$. According to Theorem \ref{thm:con_outage_exponent}, therefore, the outage probability of the user $u_{m}$ for a given $\kappa$ is given by
\begin{equation}
\begin{aligned}
	& \binom{L}{\kappa}p_{m}^{\mathrm{cop}}\left(R_{m}\left|\mathcal{D}_{\widetilde{K}_{m},L-\kappa}\right.\right)p_{\mathrm{s}}^{\kappa}q_{\mathrm{s}}^{ML-\kappa}\\
	= & \frac{L!p_{m}^{\mathrm{cop}}\left(R_{m}\left|\mathcal{D}_{\widetilde{K}_{m},L-\kappa}\right.\right)p_{\mathrm{s}}^{\kappa}}{\kappa!\left(L-\kappa\right)!}+\mathsf{O}\left(p_{\mathrm{s}}^{L}\right).
\end{aligned}
\end{equation}
Hence, according to the law of total probability, the first order approximation of the user $u_{m}$ is given by Eq. \eqref{eq:outage_RB_high}.
\end{IEEEproof}

\section{Proof of Theorem \ref{thm:outage_RB_LowSNR}}\label{app:outage_RB_LowSNR}
Consider the low SNR regime such that the sample of $\mathscr{G}\left(\mathcal{K}_{MN};\mathsf{P}\right)$ only has a few edges. For a user $u_{m}$ in the set $\mathcal{U}$, there are two cases that make $u_{m}$ unsaturated: 1) There are no $K_{m}$ non-outage RBs in $\mathcal{S}$ for $u_{m}$; and 2) There are other users competing for the same RB with $u_{m}$ and it is not saturated by the maximum $f$-matching.

In the first case, the occurrence probability of this event is given by
\begin{equation}
\begin{aligned}
	& \sum_{\kappa=L-K_{m}+1}^{L}\binom{L}{\kappa}p_{m}^{\mathrm{cop}}\left(R_{m}\left|\mathcal{D}_{K_{m},L-\kappa}\right.\right)p_{\mathrm{s}}^{\kappa}q_{\mathrm{s}}^{L-\kappa}\\
	= & p_{\mathrm{s}}^{L}+Lp_{m}^{\mathrm{cop}}\left(R_{m}\left|\mathcal{D}_{K_{m},1}\right.\right)p_{\mathrm{s}}^{L-1}q_{\mathrm{s}}+\mathsf{O}\left(q_{\mathrm{s}}\right),
\end{aligned}
\end{equation}
for $\gamma\rightarrow0$. In the second case, there will be at least two edges in the bipartite graph $\mathcal{G}\left(\mathcal{U}\cup\mathcal{S},\mathcal{E}\right)$. One is $u_{m}s_{n}$, and the other one is $u_{m'}s_{n}$. Assuming that there are only two edges in the sample of this random bipartite graph, i.e., $\mathcal{E}=\left\{u_{m}s_{n},u_{m'}s_{n}\right\}$. In the maximum $f$-matching, $u_{m}$ or $u_{m'}$ is chosen with equal probability. The outage probability of $u_{m}$ must then have the term
\begin{equation}
	\frac{1}{2}\binom{M}{2}\binom{L}{1}p_{\mathrm{s}}^{ML-2}q_{\mathrm{s}}^{2}=\mathsf{O}\left(q_{\mathrm{s}}\right),\quad\gamma\rightarrow0.
\end{equation}

Thus, if there are more than two edges in this random bipartite graph, the outage probability of $u_{m}$ must have a factor $q_{\mathrm{s}}^{x}$ with $x\geq3$. Hence, Eq. \eqref{eq:outage_RB_LowSNR} has then been established.

\section{Proof of Theorem \ref{thm:outage_chunk}}\label{app:outage_chunk}
We first consider the case of $K_{m}=1,\ldots,K_{m}^{\mathrm{th}}$. According to Lemma \ref{lem:matching_edges}, if the user $u_{m}$ is not saturated by the generated maximum $f$-matching, there are at least $L-K_{m}+1$ chunks in outage state. Let $\kappa$ be the number of outage chunks. Since the first order approximation is considered in this paper, $\kappa$ must be in the range from $L-K_{m}+1$ to $L$. According to the proof of Lemma \ref{lem:matching_edges}, the $L-K_{m}+1$ outage chunks are only outage for the user $u_{m}$. In this specific condition, the outage probability of the user $u_{m}$ is given by
\begin{equation}
\begin{aligned}
	& \binom{L}{L-K_{m}+1}p_{m}^{\mathrm{cop}}\left(R_{m}\left|\mathcal{D}_{K_{m},K_{m}-1}\right.\right)p_{\mathrm{s}}^{L-K_{m}+1}\cdot\\
	& q_{\mathrm{s}}^{ML-L+K_{m}-1}=\frac{L!p_{m}^{\mathrm{cop}}\left(R_{m}\left|\mathcal{D}_{K_{m},K_{m}-1}\right.\right)p_{\mathrm{s}}^{L-K_{m}+1}}{\left(L-K_{m}+1\right)!\left(K_{m}-1\right)!}\\
	& +\mathsf{O}\left(p_{\mathrm{s}}^{L}\right),
\end{aligned}
\end{equation}
as $\gamma\rightarrow\infty$, where the lowest order of $p_{\mathrm{s}}$ is $K_{m}-1+L-K_{m}+1=L$. For $\kappa=L-K_{m}+2,\ldots,L$, the $\kappa$ outage chunks must also be in outage only for the user $u_{m}$. If this is not true, for example, $\Delta_{1}$ outage chunks for $u_{m_{1}}$ and $\Delta_{2}$ outage chunks for $u_{m_{2}}$ with $\kappa=\Delta_{1}+\Delta_{2}$, the approximation with the lowest order of the outage probability for the user $u_{m_{i}},\,i=1,2$ is given by
\begin{equation}
\begin{aligned}
	& \binom{L}{\Delta_{1}}\binom{L-\Delta_{1}}{\Delta_{2}}p_{m}^{\mathrm{cop}}\left(R_{m}\left|\mathcal{D}_{K_{m},L-\Delta_{i}}\right.\right)p_{\mathrm{s}}^{\kappa}q_{\mathrm{s}}^{ML-\kappa}\\
	& = \frac{L!p_{m}^{\mathrm{cop}}\left(R_{m}\left|\mathcal{D}_{K_{m},L-\Delta_{i}}\right.\right)p_{\mathrm{s}}^{\kappa}}{\Delta_{1}!\Delta_{2}!\left(L-\Delta_{1}-\Delta_{2}\right)!}+\mathsf{O}\left(p_{\mathrm{s}}^{L-\Delta_{i}+\kappa}\right).
\end{aligned}
\end{equation}
Therefore, the lowest order of $p_{\mathrm{s}}$ is given by $L-\Delta_{i}+\kappa>L$. In other words, this situation results in a higher order term. Hence, according to the law of total probability, the first order approximation of the user $u_{m}$ is given by Eq. \eqref{eq:outage_chunk_1}.

Consider now the case of $K_{m}=K_{m}^{\mathrm{th}}+1,\ldots,\widetilde{K}_{m}$. According to Lemma \ref{lem:matching_edges}, if the user $u_{m}$ is not saturated by the generated maximum $f$-matching, there are at least $ML-M\left(K^{\mathrm{sum}}-1\right)=M\left(L-K^{\mathrm{sum}}+1\right)$ chunks in outage state. According to the proof of Lemma \ref{lem:matching_edges}, the only outage user $u_{m}$ has $K_{m}-1$ non-outage chunks. Therefore, the first order approximation of the outage probability for the user $u_{m}$ is given by
\begin{equation}
\begin{aligned}
	& \frac{\binom{L}{K^{\mathrm{sum}}-1}}{M}p_{m}^{\mathrm{cop}}\left(R_{m}\left|\mathcal{D}_{K_{m},K_{m}-1}\right.\right)p_{\mathrm{s}}^{M\left(L-K^{\mathrm{sum}}+1\right)}q_{\mathrm{s}}^{M\left(K^{\mathrm{sum}}-1\right)}\\
   	& =\frac{L!p_{m}^{\mathrm{cop}}\left(R_{m}\left|\mathcal{D}_{K_{m},K_{m}-1}\right.\right)p_{\mathrm{s}}^{M\left(L-K^{\mathrm{sum}}+1\right)}}{M\left(L-K^{\mathrm{sum}}+1\right)!\left(K^{\mathrm{sum}}-1\right)!}\\
   	& +\mathsf{O}\left(p_{\mathrm{s}}^{M\left(L-K^{\mathrm{sum}}+1\right)+K_{m}-1}\right),
\end{aligned}
\end{equation}
when $\gamma\rightarrow\infty$.

If it happens that $K_{m}=K_{m}^{\mathrm{th}}$ and $\left(M-1\right)|MK_{m}^{\mathrm{sum}}$, the outage probability of the user $u_{m}$ is clearly the summation of Eq. \eqref{eq:outage_chunk_1} and Eq. \eqref{eq:outage_chunk_2}.

\section{Proof of Lemma \ref{lem:perfect_f_matching}}\label{app:perfect_f_matching}
Suppose $\mathcal{G}(\mathcal{\widetilde{U}}\cup\mathcal{S},\widetilde{\mathcal{E}})$ has a perfect matching $\widetilde{\mathcal{M}}$. For each edge $u_{m}s_{n}\in\mathcal{E}$, let $l\left(u_{m}s_{n}\right)$ denote the number of edges of $\widetilde{\mathcal{M}}$ connecting $\mathcal{X}_{u_{m}}$ and $s_{n}\in\mathcal{S}$. Then the number of edges of $\widetilde{\mathcal{M}}$ incident with $\mathcal{X}_{u_{m}}$ is exactly $\sum_{s_{n}\in\mathcal{S}}l\left(u_{m}s_{n}\right)$, which is also equal to $\left|\mathcal{X}_{u_{m}}\right|=K_{m}$. Since $\widetilde{\mathcal{M}}$ is a perfect matching, then
\begin{equation}
	|\widetilde{\mathcal{M}}|=\sum_{u_{m}\in\mathcal{U}}\sum_{s_{n}\in\mathcal{S}}l\left(u_{m}s_{n}\right)=MN
\end{equation}
is the maximum number of edges in any matching of $\mathcal{G}(\mathcal{\widetilde{U}}\cup\mathcal{S},\widetilde{\mathcal{E}})$. This fact shows that if $\mathcal{X}_{u_{m}}$ is seen as $u_{m}$, this is a perfect $f$-matching $\mathcal{M}_{f}^{\mathrm{p}}$. If this is not true, there will be more edges in $\mathcal{M}_{f}^{\mathrm{p}}$ than in $\widetilde{\mathcal{M}}$. In fact, when constructing $\mathcal{G}(\mathcal{\widetilde{U}}\cup\mathcal{S},\widetilde{\mathcal{E}})$, we can dispatch one edge incident with $u_{m}$ in $\mathcal{M}_{f}^{\mathrm{p}}$ to one vertex in $\mathcal{X}_{u_{m}}$. As a result, a new matching which contains more edges than $\widetilde{\mathcal{M}}$ can be obtained. This contradicts the fact that $\widetilde{\mathcal{M}}$ is a perfect matching of $\mathcal{G}(\mathcal{\widetilde{U}}\cup\mathcal{S},\widetilde{\mathcal{E}})$.

\ifCLASSOPTIONcaptionsoff
  \newpage
\fi



\bibliographystyle{IEEEtran}
\bibliography{../../../BibTex/IEEEabrv,../../../BibTex/library}
\end{document}